\documentclass[10pt,authoryear]{article}
\usepackage{appendix}
\usepackage{setspace}
\usepackage[font=small,  margin=2cm]{caption}
\usepackage[tbtags]{amsmath}
\usepackage{amsthm,amssymb,amsfonts}
\usepackage{natbib}
\usepackage{multirow}
\usepackage{lscape}
\usepackage{subfigure}
\usepackage{makecell}
\usepackage{exscale}
\usepackage{booktabs}
\usepackage{array}
\usepackage{fullpage}
\usepackage{url}
\usepackage{algorithm}
\usepackage{algpseudocode}
\usepackage{bm}
\usepackage{smile}
\usepackage{mathtools}
\usepackage{wrapfig}
\usepackage{lipsum}
\usepackage{threeparttable}
\usepackage{mathrsfs}
\usepackage{relsize}
\usepackage{dsfont}
\usepackage{multirow}
\usepackage{apalike}
\usepackage[usenames,dvipsnames,svgnames,table]{xcolor}
\usepackage[colorlinks=true,linkcolor=blue,urlcolor=blue,citecolor=blue]{hyperref}

\newcommand{\bargmin}{\mathop{\mathrm{arg\ min}}}
\numberwithin{equation}{section}
\numberwithin{theorem}{section}
\numberwithin{corollary}{section}
\numberwithin{definition}{section}

\begin{document}

\title{\LARGE Matrix Factor Analysis: From Least Squares to Iterative Projection}

	\author{
	Yong He
		\footnotemark[1],
	Xinbing Kong\footnotemark[2],
Long Yu\footnotemark[3],
	Xinsheng Zhang
		\footnotemark[4],
Changwei Zhao\footnotemark[1]
	}
\renewcommand{\thefootnote}{\fnsymbol{footnote}}
\footnotetext[1]{Institute of Financial Studies, Shandong University, China; e-mail:{\tt heyong@sdu.edu.cn, zhaochangwei@mail.sdu.edu.cn}}
\footnotetext[2]{Nanjing Audit University, China; e-mail:{\tt xinbingkong@126.com }}
\footnotetext[3]{National University of Singapore, Singapore; email:{\tt loyu@umich.edu}.}
\footnotetext[4]{Department of Statistics, School of Management at Fudan University, China; e-mail:{\tt xszhang@fudan.edu.cn }}

\maketitle

In this article, we study large-dimensional matrix factor models and estimate the factor loading matrices and factor score matrix by minimizing square loss function. Interestingly, the resultant estimators coincide with the Projected Estimators (PE) in \cite{Yu2021Projected}, which was proposed from the perspective of simultaneous reduction of the dimensionality and the magnitudes of the idiosyncratic error matrix. In other word, we provide a least-square interpretation of the PE for matrix factor model, which parallels to the least-square interpretation of the PCA for the vector factor model. We derive the convergence rates of the theoretical minimizers under sub-Gaussian tails. Considering the robustness to the heavy tails of the idiosyncratic errors, we extend the least squares to minimizing the Huber loss function, which leads to a weighted iterative projection approach to compute and learn the parameters.  We also derive the convergence rates of the theoretical minimizers of the Huber loss function under bounded $(2+\epsilon)$th moment of the idiosyncratic errors. We conduct extensive numerical studies to investigate the empirical performance of the proposed Huber estimators relative to the state-of-the-art ones. The Huber estimators perform robustly and much better than existing ones when the data are heavy-tailed, and as a result can be used as a safe replacement in practice. An application to a Fama-French financial portfolio dataset demonstrates the empirical advantage of the Huber estimator.
\vspace{2em}


\section{Introduction}
Large-dimensional factor model has been a powerful tool of summarizing information from large datasets and draws growing attention in the era of ``big-data"  when more and more records of variables are available. The last two decades have seen many studies on large-dimensional approximate vector factor models, since the seminal work by \cite{bai2002determining} and \cite{stock2002forecasting},
see for example, the representative works by \cite{bai2003inferential},\cite{onatski09}, \cite{ahn2013eigenvalue}, \cite{fan2013large}, and \cite{Trapani2018A}, \cite{Sahalia2017Using}, \cite{Sahalia2020High}. These works all require the fourth moments (or even higher moments) of factors and idiosyncratic errors, and there are some works on relaxing the restrictive moment conditions, see the endeavors by \cite{yu2019robust}, \cite{Chen2021Quantile} and \cite{He2020large}.

In the last few years, large-dimensional matrix factor models have drawn much attention in view of the fact that observations are usually well structured to be an array, such as in macroeconomics and finance, see \citet{fan2021} for further examples of matrix observations. The seminal work is the one by \cite{wang2019factor}, who proposed the following formulation for matrix time series observations $\{\Xb_t, 1\leq t\leq T\}$:
\begin{equation}\label{equ:matrixfactormodel}
  \underbrace{\Xb_t}_{p_1\times p_2}=\underbrace{\Rb_0}_{p_1\times k_1}\times \underbrace{\Fb_{0t}}_{k_1\times k_2}\times \underbrace{\Cb^\top_0}_{k_2\times p_2}+  \underbrace{\Eb_t}_{p_1\times p_2},
\end{equation}
where $\Rb_0$ is the row factor loading matrix exploiting the variations of $\Xb_{t}$ across the rows, $\Cb_0$ is the $p_{2}\times k_{2}$ column factor loading matrix reflecting the differences across the columns of $\Xb_{t}$, $\Fb_{0t}$ is the common factor matrix for all cells in $\Xb_{t}$ and $\Eb_{t}$ is the idiosyncratic component. \citet{wang2019factor} proposed estimators of the factor loading matrices and numbers of the row and column factors based on an eigen-analysis of the auto-cross-covariance matrix.
\citet{fan2021} proposed an $\alpha$-PCA method for inference of (\ref{equ:matrixfactormodel}), which conducts eigen-analysis of a weighted average of the sample mean and the column (row) sample covariance matrix; \citet{Yu2021Projected} proposed a projected estimation method which further improved the estimation efficiency of the factor loading matrices and the numbers of factors. \cite{He2021Vector} proposed a strong rule to determine whether there is a factor structure of  matrix time series and a sequential procedure to determine the numbers of factors. Extensions and applications of the matrix factor model include the dynamic transport network in the international trade flows by \cite{Chen2020Modeling}, the constrained matrix factor model by \citet{chen2019constrained},  the threshold matrix factor model in \citet{liu2019helping} and the online change point model by \cite{he2021online}. \cite{Gao2021A} proposed an interesting two-way factor model and provided solid theory. \cite{Jing2021Community} introduced mixture multilayer stochastic block model and proposed
a tensor-based algorithm (TWIST) to reveal both global/local memberships
of nodes, and memberships of layers for worldwide trading networks. There exist some recent works in the broader context of tensor factor model, see for example, \cite{han2020rank,chen2020semiparametric,Han2020Tensor,lam2021rank,tensorTS,Chang2021Modelling,chenlam2022,chen2022factor}. In general, the projection-based method would lead to more accurate estimation of the loading spaces, but would be computationally a bit demanding \citep{Yu2021Projected,Han2020Tensor}.

A natural competitor of (\ref{equ:matrixfactormodel})
is  the group (vector) factor model (see e.g. \citealp{ando}), however, in such a model  only one
cross-section exists, which contains variables of the same
nature well-grouped with known or unknown group membership.
The model organizes the common factors into groups, and characterizes the interrelations within
and between groups  by such group factors. In contrast, the data $%
\Xb_{t}$ in matrix factor model (\ref{equ:matrixfactormodel}) are genuinely matrix-valued, with two cross-sectional
dimensions of different nature.
The common components $\Rb_0 \Fb_{0t}\Cb_0 ^\top $ in (\ref{equ:matrixfactormodel}) reflect the
interplay between the two different cross-sections. In the
context of recommending systems illustrated in \cite{He2021Vector}, ratings are high whenever the purchasers'
consumption preferences{\ (rows in $\Rb_0$)} match the underlying
characteristics of items displayed online{\ (rows in $\Cb_0$), thus the matrix factor model
naturally captures the interactive effect between the row and column
cross sections.} In this context, matrix factor model (\ref{equ:matrixfactormodel}) takes the intrinsic matrix nature of the data into account and shows advantage  over  models based on vectorising $\Xb_{t}$, where the presence of groups
arises from artificially stacking the columns (rows) of matrix-valued data.
Moreover, the matrix factor model in (\ref{equ:matrixfactormodel})  has a more parsimonious factor structure, thus being
statistically and computationally more efficient \citep{fan2021,He2021Vector,Yu2021Projected}.

\begin{figure}[!h]
  \centering
  \begin{minipage}[!t]{0.48\linewidth}
    \includegraphics[width=8cm]{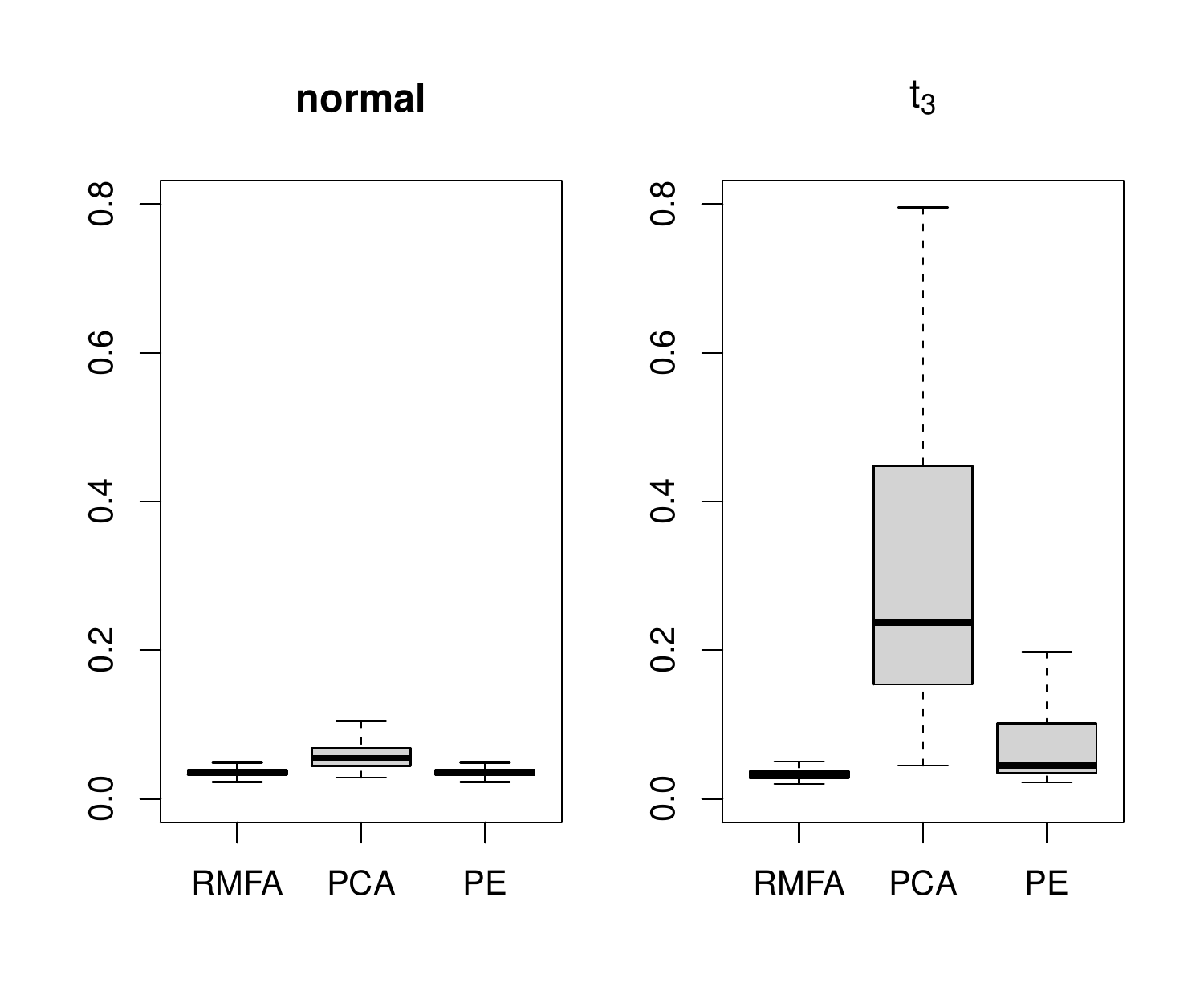}\\
  \end{minipage}
  \begin{minipage}[!t]{0.48\linewidth}
    \includegraphics[width=8cm]{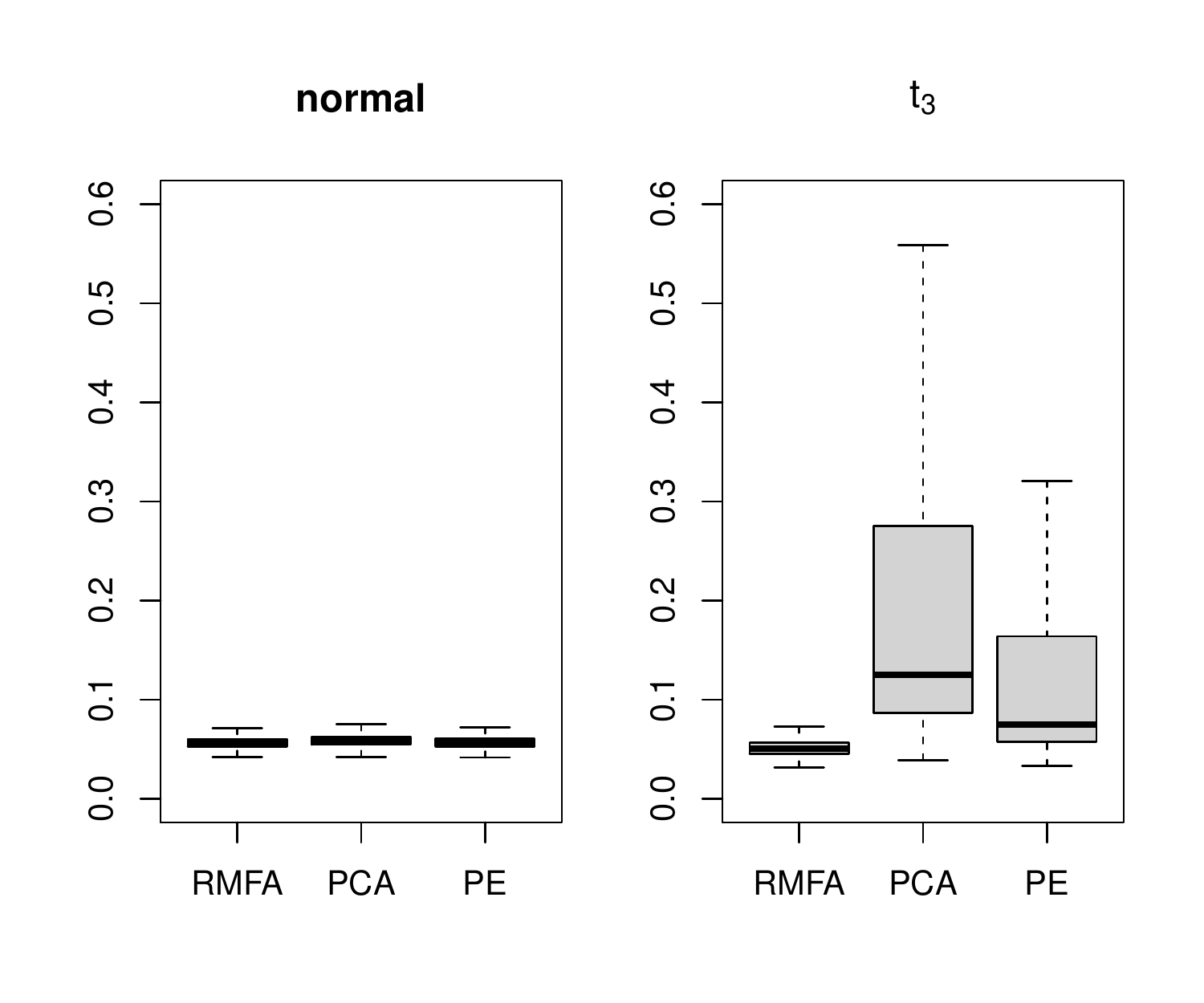}\\
  \end{minipage}
 	\caption{Boxplot of the distance between the estimated loading space and the true loading space by RMFA, PCA and PE methods under different distributions (matrix normal and matrix $t_3$). $p_1=20,p_2=T=50$. The left two plots depict the distances between the estimated loading space $\hat\Rb$ and true loading space $\Rb_0$, and the right two plots depict the distances between the estimated loading space $\hat \Cb$ and true loading space $\Cb_0$.}\label{fig:1}
 \end{figure}

In the current work, we find the equivalence between the least square approach and the PE method by \cite{Yu2021Projected}.  In other word, we provide the least squares interpretation
of the PE for matrix factor model, which parallels to the least squares interpretation of traditional PCA
for vector factor models. This finding provides another rationale for the PE method, which was initially proposed for reducing the magnitudes of the idiosyncratic error components and thereby increasing the signal-to-noise ratio. Motivated by the least squares formulation, we further propose a robust method for estimating large-dimensional matrix factor models, by substituting the least squares loss function with the Huber Loss function. The resultant estimators of factor loading matrices can be simply viewed as the eigenvectors of weighted sample covariance matrices of the projected data, which are easily obtained by an iterative algorithm. As far as we know, this is the first methodology work, with weighted iterative projection algorithms, on robust analysis of matrix factor models. As an illustration, we check the sensitivity of the $\alpha$-PCA method by \cite{fan2021} and the Projected Estimation (PE) method (or least squares minimization method) in \cite{Yu2021Projected} to the heavy-tailedness of the idiosyncratic errors with a synthetic dataset. We generate the idiosyncratic errors from matrix-variate normal, matrix-variate  $t_3$  distributions that will be described in detail in  Section \ref{sec:5}. Figure \ref{fig:1} depicts the boxplots of the estimation errors of the factor loading matrices over 1000 replications. It is  clearly seen that the $\alpha$-PCA and PE methods lead to much bigger biases and higher dispersions as the distribution tails become heavier. The proposed Robust Matrix Factor Analysis (RMFA)  method still works quite satisfactorily when the idiosyncratic errors are from matrix-variate  $t_3$ distribution.

To do factor analysis, the first step is to determine the number of factors. As for the Matrix Factor Model (MFM), both the row and column factor numbers should be determined in advance. \cite{wang2019factor} proposed an estimation method based on ratios of consecutive eigenvalues of auto-covariance matrices; \cite{fan2021} proposed an $\alpha$-PCA based eigenvalue-ratio method and \cite{Yu2021Projected} further proposed a projection-based iterative eigenvalue-ratio method. All the methods above borrowed idea from \cite{ahn2013eigenvalue} and, to the best of our knowledge, \cite{He2021Vector} is the only work that determines the factor numbers of MFM from the perspective of sequential hypothesis testing and the authors also provide a strong rule to determine whether there is a row/column factor structure in the matrix time series. However, none methods mentioned above take the well-known heavy-tailedness of the data into account (see also Figure \ref{fig:kurtosis} in the real data section). In the current work, we also present a robust iterative eigenvalue-ratio method to estimate the numbers of factors following the Huber \citep{Huber1964Robust} loss formulation.

The contributions of the present work lie in the following aspects. Firstly, we formulate the estimation of MFM from the least squares point of view and find that minimizing the square loss function under the identifiability conditions naturally leads to the iterative projection method invented in \cite{Yu2021Projected} and thus enjoys the nice properties of the PE method. Secondly, we further propose a robust estimation method for MFM by taking the Huber loss in place of the least squares loss. We also propose an iterative weighted projection approach to solve the corresponding optimization problem, which in each iteration is doable by simple PCA solution. Thirdly, we propose an iterative algorithm for estimating the row/column factor number robustly.  Lastly, we derive the convergence rates of the theoretical minimizers, first time for the matrix factor model. To the best of our knowledge, this is the first work that derives the convergence rates of the theoretical minimizers of Huber loss function under bounded $(2+\epsilon)$th moment condition of the idiosyncratic errors for the matrix factor model, with the theoretical tool of self-normalized large/moderate deviations \citep{shao1997self,jing2008towards}.

The rest of the article goes as follows. In Section 2, we first formulate the estimation of factor loading matrices and factor score matrix by minimizing the square loss function and provide solutions to the optimization problem, from which we can see its equivalence to the projected estimation method. In Section 3, we investigate the theoretical minimizers of the least squares   under mild  conditions.  In Section 4,  we provide robust estimators by considering the Huber loss function and present detailed algorithm to obtain the minimizers. We also propose robust estimators for the pair of factor numbers. In Section 5, we conduct thorough numerical studies to illustrate the advantages of the RMFA method and the robust iterative eigenvalue-ratio method  over the state-of-the-art methods. In Section 6, we analyze a financial dataset to illustrate the empirical usefulness of the proposed methods.
 We discuss possible future research direction and conclude the article in Section 7. The proofs of the main theorems and
additional details are collected in the supplementary materials.

Before ending this section, we introduce the notations used throughout the paper. For any vector $\bmu=(\mu_1,\ldots,\mu_p)^\top \in \RR^p$, let $\|\bmu\|_2=(\sum_{i=1}^p\mu_i^2)^{1/2}$, $\|\bmu\|_\infty=\max_i|\mu_i|$. For a real number $a$, denote  $[a]$ as the largest integer smaller than or equal to $a$,  let $sgn(a) = 1$ if $a\geq 0$ and
$sgn(a) =-1$ if $a<0$. For a square matrix $\Ab$ whose $j$th diagonal element is denoted as
$A_{jj}$, define $sgn(\Ab)$ as a diagonal matrix whose $j$th diagonal element is equal to $sgn(A_{jj})$. Let $I(\cdot)$ be the indicator function and ${\rm diag}(a_1,\ldots,a_p)$ be a $p\times p$ diagonal matrix, whose diagonal entries are $a_1\ldots,a_p$.   For a matrix $\Ab$, let $\mathrm{A}_{ij}$ (or $\mathrm{A}_{i,j}$) be the $(i,j)$-th entry of $\Ab$, $\Ab^\top$ the transpose of $\Ab$, ${\rm Tr}(\Ab)$ the trace of $\Ab$, $\text{rank}(\Ab)$ the rank of $\Ab$ and $\text{diag}(\Ab)$ a vector composed of the diagonal elements of $\Ab$. Denote $\lambda_j(\Ab)$ as the $j$-th largest eigenvalue of a nonnegative definitive matrix $\Ab$, and let $\|\Ab\|$ be the spectral norm of matrix $\Ab$ and $\|\Ab\|_F$ be the Frobenius norm of $\Ab$. For two series of random variables, $X_n$ and $Y_n$, $X_n\asymp Y_n$ means $X_n=O_p(Y_n)$ and $Y_n=O_p(X_n)$. For two random variables (vectors) $\bX$ and $\bY$, $\bX\stackrel{d}{=}\bY$ means the distributions of $\bX$ and $\bY$ are identical. The constants $c,C_1,C_2$ in different lines can be nonidentical.

	\section{Least Squares and Projected Estimation}
	In this section, we establish the equivalence between minimizing the square loss function under the identifiability conditions and the projection estimation procedure. We show that both angles coincide in the same iterative algorithm. We assume the matrix factor model (\ref{equ:matrixfactormodel}) and let $\Sbb_t=\Rb_0 \Fb_{0t} \Cb_0^\top$ be the common component matrix. The loading matrices  $\Rb_0$  and  $\Cb_0$  in model (\ref{equ:matrixfactormodel}) are not separately identifiable. Without loss of generality, for identifiability issue, we assume that $ \Rb^\top_0 \Rb_0/p_1 = \Ib_{k_1} $ and $ \Cb^\top_0 \Cb_0/p_2 = \Ib_{k_2}$, see also \cite{fan2021,Yu2021Projected}.
	
	From (\ref{equ:matrixfactormodel}), it is a natural idea to estimate $\mathbf{R}_0$ and $\mathbf{C}_0$ by minimizing the square loss under the identifiability condition:
	
	\begin{equation} \label{LOS}
\begin{array}{cc}
		 &\min_{\{\mathbf{R},\mathbf{C},\mathbf{F}_t\}}L_1(\mathbf{R},\mathbf{C},\mathbf{F}_t)=\dfrac{1}{T}\sum_{t=1}^{T}\Vert \mathbf{X}_t - \mathbf{R} \mathbf{F}_t \mathbf{C}^\top \Vert ^2 _F, \\
&\text{s.t.}\dfrac{1}{p_1} \mathbf{R}^\top \mathbf{R} = \mathbf{I}_{k_1},\dfrac{1}{p_2} \mathbf{C}^\top \mathbf{C} = \mathbf{I}_{k_2}.
\end{array}
	\end{equation}
The right hand side of (\ref{LOS}) can be simplified as:
	\[\dfrac{1}{T}\sum_{t=1}^{T}\Vert \mathbf{X}_t - \mathbf{R} \mathbf{F}_t \mathbf{C}^\top \Vert ^2 _F = \dfrac{1}{T}\sum_{t=1}^{T}\left[\text{Tr}(\mathbf{X_t}^\top\mathbf{X_t})-2\text{Tr}(\mathbf{X_t}^\top \mathbf{R} \mathbf{F_t} \mathbf{C}^\top) +p_1 p_2 \text{Tr}(\mathbf{F_t}^\top \mathbf{F_t}) \right].
	\]
	The optimization is non-convex in $\{\mathbf{R},\mathbf{C},\mathbf{F}_t\}$, but given the others, the loss function is convex over the remaining parameter. For instance, given $(\mathbf{R},\mathbf{F}_t)$, $L_1(\mathbf{R},\mathbf{C},\mathbf{F}_t)$ is convex over $\Cb$.
	Then we first assume that $(\Rb,\Cb)$ are given and  solve the optimization problem on $\mathbf{F_t}$. For each $t$, taking ${\partial L_{1}(\mathbf{R},\mathbf{C})}/{\partial \mathbf{F_t}} = 0$, we obtain
	\[ \mathbf{F_t}=\dfrac{1}{p_1p_2}\mathbf{R}^\top \mathbf{X_t} \mathbf{C}.
	\]
	Thus by substituting $\mathbf{F_t}=\mathbf{R}^\top \mathbf{X_t} \mathbf{C}/({p_1p_2})$ in the loss function $L_1(\mathbf{R},\mathbf{C},\mathbf{F}_t)$, we further have
	\begin{equation} \label{LLOS}
		\begin{array}{cc} \min_{\{\mathbf{R},\mathbf{C}\}}L_{1}(\mathbf{R},\mathbf{C})=\dfrac{1}{T}\sum_{t=1}^{T}\left[\text{Tr}(\mathbf{X_t}^\top\mathbf{X_t})-\dfrac{1}{p_1p_2}\text{Tr}(\mathbf{X_t}^\top \mathbf{R} \mathbf{R}^\top \mathbf{X_t} \mathbf{C} \mathbf{C}^\top) \right],\vspace{2ex} \\
\text{s.t.}\dfrac{1}{p_1} \mathbf{R}^\top \mathbf{R} = \mathbf{I}_{k_1},\dfrac{1}{p_2} \mathbf{C}^\top \mathbf{C} = \mathbf{I}_{k_2}.
\end{array}
	\end{equation}
	The Lagrangian function is as follows:
	 \[\min_{\{\mathbf{R},\mathbf{C}\}}\mathcal{L}_{1}=L_{1}(\mathbf{R},\mathbf{C}) + \text{Tr} \left[\mathbf{\Theta} (\dfrac{1}{p_1} \mathbf{R}^\top \mathbf{R} - \mathbf{I}_{k_1})\right] + \text{Tr} \left[\mathbf{\Lambda} (\dfrac{1}{p_2} \mathbf{C}^\top \mathbf{C} - \mathbf{I}_{k_2})\right],
	\]
	where the Lagrangian multipliers $\mathbf{\Theta}$ and $\mathbf{\Lambda}$ are symmetric matrices.
	According to the KKT condition, let
	\[\dfrac{\partial \mathcal{L}_{1}}{\partial \mathbf{R}} = -\dfrac{1}{T} \sum_{t=1}^{T} \dfrac{2}{p_1p_2}\mathbf{X}_t \mathbf{C} \mathbf{C}^\top \mathbf{X}_t ^\top \mathbf{R} + \dfrac{2}{p_1} \mathbf{R} \mathbf{\Theta} =0,
	\]
	\[\dfrac{\partial \mathcal{L}_{1}}{\partial \mathbf{C}} = -\dfrac{1}{T} \sum_{t=1}^{T} \dfrac{2}{p_1p_2}\mathbf{X}_t^\top \mathbf{R} \mathbf{R}^\top \mathbf{X}_t \mathbf{C} + \dfrac{2}{p_2} \mathbf{C} \mathbf{\Lambda} =0,
	\]
	respectively, it holds that
	\begin{equation}\label{equ:leasteigen}
	\left\{\begin{array}{ccc}
	\left(\dfrac{1}{Tp_2} \sum\limits_{t=1}^{T} \mathbf{X}_t \mathbf{C} \mathbf{C}^\top \mathbf{X}_t ^\top\right) \mathbf{R}   = \mathbf{R} \mathbf{\Theta},\vspace{2ex} \\
    	\left(\dfrac{1}{Tp_1} \sum\limits_{t=1}^{T} \mathbf{X}_t^\top \mathbf{R} \mathbf{R}^\top \mathbf{X}_t\right) \mathbf{C}   = \mathbf{C} \mathbf{\Lambda},
    \end{array}\right. \text{or} \ \ \left\{\begin{array}{ccc}
	\Mb_c \mathbf{R}   = \mathbf{R} \mathbf{\Theta},\vspace{2ex} \\
    \Mb_r 	 \mathbf{C}   = \mathbf{C} \mathbf{\Lambda},
    \end{array}\right.
    \end{equation}
where
\[
\Mb_c=\dfrac{1}{Tp_2} \sum_{t=1}^{T} \mathbf{X}_t \mathbf{C} \mathbf{C}^\top \mathbf{X}_t ^\top, \ \ \  \Mb_r=\dfrac{1}{Tp_1} \sum_{t=1}^{T} \mathbf{X}_t^\top \mathbf{R} \mathbf{R}^\top \mathbf{X}_t .
\]
We denote the first $k_1$ eigenvectors of $\Mb_c$ as $\{\br(1),\ldots,\br(k_1)\}$ and the corresponding eigenvalues as $\{\theta_1,\ldots,\theta_{k_1}\}$. Similarly, we denote the first $k_2$ eigenvectors of $\Mb_r$ as $\{\bc(1),\ldots,\bc(k_2)\}$ and the corresponding eigenvalues as $\{\lambda_1,\ldots,\lambda_{k_2}\}$.
    From (\ref{equ:leasteigen}),   $\Rb=\sqrt{p_1}(\br(1),\ldots,\br(k_1))$, $\Cb=\sqrt{p_2}(\bc(1),\ldots,\bc(k_2))$, $\bTheta=\text{diag}(\theta_1,\ldots,\theta_{k_1})$, $\bLambda=\text{diag}(\lambda_1,\ldots,\lambda_{k_2})$ satisfy the KKT condition. However, $\Mb_c$ relies on the unknown column factor loading $\Cb$ while  $\Mb_r$ relies on the unknown row factor loading $\Rb$, which motivates us to consider an iterative procedure to get the estimators, where in each iteration a simple PCA manipulation is enough. This turns out to be the same as our projected estimation procedure in \cite{Yu2021Projected}.
We summarized the algorithm in Algorithm \ref{alg:project}, where the initial estimators $\hat \Rb^{(0)}$ and $\hat\Cb^{(0)}$ could be selected as the $\alpha$-PCA estimators.

	\begin{algorithm}[!h]
	\caption{Least squares (Projection Estimation) for estimating matrix factor spaces}\label{alg:project}
	{\bf Input:} Data matrices $\{\Xb_t\}_{t\le T}$, the pair of row and column factor numbers $k_1$ and $k_2$\\
	{\bf Output:} Factor loading matrices $\tilde\Rb$ and $\tilde\Cb$
	\begin{algorithmic}[1]
		\State  obtain the  initial estimators $\hat \Rb^{(0)}$ and $\hat\Cb^{(0)}$ by $\alpha$-PCA with $\alpha=0$;
		
		\State project the data matrices to lower dimensions by defining: $\hat\Yb_t=\Xb_t\hat\Cb^{(0)}$ and $\hat\Zb_t=\Xb_t^\top\hat\Rb^{(0)}$;
		
		\State given $\hat \Yb_t$ and $\hat\Zb_t$, define $\hat\Mb_r=(Tp_1)^{-1}\sum_{t=1}^{T}\hat\Yb_t\hat\Yb_t^\top$ and $\hat\Mb_c=(Tp_2)^{-1}\sum_{t=1}^{T}\hat\Zb_t\hat\Zb_t^\top$, and obtain the the leading $k_1$ eigenvectors of $\hat\Mb_c$, denote as $\{\hat\br(1),\ldots,\hat\br(k_1)\}$ and the the leading $k_2$ eigenvectors of $\hat\Mb_r$, denoted as $\{\hat\bc(1),\ldots,\hat\bc(k_2)\}$; Then update $\hat\Rb$ and $\hat\Cb$ as $\hat\Rb^{(1)}=\sqrt{p_1}(\hat\br(1),\ldots,\hat\br(k_1))$ and $\hat\Cb^{(1)}=\sqrt{p_2}(\hat\bc(1),\ldots,\hat\bc(k_2))$.
\State repeat step 2 and 3 until convergence and output the estimators from the last step and denote them as $\tilde \Rb$ and $\tilde \Cb$.
		
	\end{algorithmic}
\end{algorithm}

In the following, let's briefly review the Projection Estimation (PE) method in \cite{Yu2021Projected}.
  First assume $\Cb$ is known and satisfies the orthogonal condition $\Cb^\top\Cb/p_2=\Ib_{k_2}$. In \cite{Yu2021Projected}, we  projected the data matrix to a lower dimensional space by setting

\begin{equation}\label{equ:Yu}
\Yb_t=\frac{1}{p_2}\Xb_t\Cb=\frac{1}{p_2}\Rb\Fb_t\Cb^\top\Cb+\frac{1}{p_2}\Eb_t\Cb:=\Rb\Fb_t+\tilde\Eb_t.
\end{equation}

After projection, $\Yb_t$ lies in a much lower column space than $\Xb_t$, and $\Fb_t $ and $\tilde\Eb_t$ are deemed as factors and idiosyncratic errors for $\Yb_t$, respectively.  For the $i$th row of $\tilde\Eb_t$, denoted as $\tilde\be_{t,i\cdot}$, $\mathbb{E}\|\tilde\be_{t,i\cdot}\|^2\le cp_2^{-1}$  as long as the original errors $\{e_{t,ij}\}^{p_2}_{j=1}$ are weakly dependent column-wise. Thus $\Yb_t$ can be viewed as satisfying a nearly noise-free factor model when $p_2$ is large.  In other word, projecting the observation
matrix onto the column factor space would not only simplify factor analysis for matrix sequences
to that of a lower-dimensional tensor, but also reduce the magnitudes of the
idiosyncratic error components, thereby increasing the signal-to-noise ratio. The next step of the PE method is to construct the row sample covariance matrix with $\{\Yb_t\}$, which is exactly $\Mb_c$ up to a constant. Once $\Mb_c$ is constructed, the following steps of the PE method are exactly the same as stated in Algorithm \ref{alg:project}.
Surprisingly, we find that two totally different angles, least squares and projection approach coincide in matrix factor analysis.

	\section{Theoretical results}\label{sec:4}
 Though not computationally reachable, we establish the convergence rate of the theoretical minimizers of (\ref{LOS}) under sub-Gaussian tails of idiosyncratic errors, instead of the two-step iteration estimators in Algorithm 1 by \cite{Yu2021Projected}.
     For optimization problem (\ref{LOS}), let $\theta=(\Rb,\Fb_1,\cdots,\Fb_T,\Cb)$  and $\theta_0=(\Rb_0,\Fb_{01},\cdots,\Fb_{0T},\Cb_0)$ be the true parameters, where $\Rb = (\rb_1, \cdots, \rb_{p_1})^{\top}, \Cb = (\cbb_1, \cdots, \cbb_{p_2})^{\top}$.  In this section, we present the asymptotic properties of the theoretical minimizers $\hat{\theta}$, defined as
    $$\hat{\theta} = (\hat{\Rb},\hat{\Fb}_1,\cdots,\hat{\Fb}_T,\hat{\Cb}) = \bargmin_{\theta\in\Theta}\dfrac{1}{T}\sum_{t=1}^{T}\Vert \mathbf{X}_t - \mathbf{R} \mathbf{F}_t \mathbf{C}^\top \Vert ^2 _F,$$
	where
	\begin{eqnarray*}
	\Theta=\Big\{ \theta\text{ : } \Rb\in \cA\subset \RR^{p_1\times k_1}, \Cb \in \cB \subset \RR^{p_2\times k_2}, \Fb_t \in \cF\subset \RR^{k_1\times k_2} \text{ for all }  t,
 \dfrac{1}{p_1} \mathbf{R}^\top \mathbf{R} = \mathbf{I}_{k_1},\dfrac{1}{p_2} \mathbf{C}^\top \mathbf{C} = \mathbf{I}_{k_2} \Big\}.
	\end{eqnarray*}

	To obtain the theoretical properties of $\hat \theta$, we assume that the following three assumptions hold.

    \vspace{0.5em}

	\textbf{Assumption 1: }$\cA, \cB$ and $\cF$ are compact sets and $\theta_0 \in \Theta$. $\lVert \Rb_0 \rVert_F/\sqrt{p_1}$ and $\lVert \Cb_0 \rVert_F /\sqrt{p_2}$ are bounded. The factor matrix satisfies
	\[
	\dfrac{1}{T} \sum\limits_{t=1}^T \Fb_{0t} \Fb_{0t}^\top \rightarrow \mathbf{\Sigma}_1 \text{ and } \dfrac{1}{T} \sum\limits_{t=1}^T \Fb_{0t}^\top \Fb_{0t} \rightarrow \mathbf{\Sigma}_2,  \ \ \text{as} \ \ T\rightarrow\infty,
	\]
	where $\mathbf{\Sigma}_i,i=1,2$ is a $k_i \times k_i$ positive diagonal matrix with bounded distinctive eigenvalues.
	
   \vspace{0.5em}

	\textbf{Assumption 2: }Given $\{ \Fb_{0t}, t=1,\cdots,T \}, \{ e_{ij,t}\}$ are independent across $i,j$ and $t$.
	
\vspace{0.5em}

	\textbf{Assumption 3: }For some $K>0$, $E(|e_{ij,t}|^p|\{\Fb_{0t}\})\leq K^p p^{p/2}$ for all $p>1$, and $E(e_{ij,t}|\{\Fb_{0t}\})=0$.

\vspace{0.5em}

Assumption 1 is standard in large factor models,  $\mathbf{\Sigma}_1,\mathbf{\Sigma}_2$ are assumed to be diagonal matrices with distinctive diagonal elements  for further identifiability issue, and we refer, for example, to \cite{bai2013principal}, \cite{fan2021} and \cite{He2021Vector}. Assumption 2  assumes that the
error terms are {\textit {i.i.d}} conditional on the matrix factors, but they may not be {\textit {i.i.d}} unconditionally.
Although this condition seems to be restrictive, it is an exchange for simplicity, see for example \cite{Chen2021Quantile}. Assumption 3 assumes that the idiosyncratic errors have sub-Gaussian tails given $\{\Fb_{0t}\}$, i.e.,  there are positive constants $C, \nu$ such that for every $t>0$, $P(|e_{ij,t}|>t|\{\Fb_{0t}\})\leq C\exp(-\nu t^2)$. The sub-Gaussian condition is for simplicity of theoretical proof, which is common in high-dimensional statistical inference, see for example \cite{Wainwright2019}.

The following theorem presents the asymptotic property of $\hat{\Rb}, \hat{\Cb} \text{ and }\hat{\Fb}_t, t=1,\cdots,T$ in terms of both Frobenius norm and spectral norm.

	\begin{theorem}\label{the1}
 Let	$\hat{\Sbb}_1 = sgn\Big(\dfrac{1}{T} \sum\limits_{t=1}^{T} (\hat{\Fb}_{t} \Fb_{0t}^\top)\Big)$ and $\hat{\Sbb}_2 = sgn\Big(\dfrac{1}{T} \sum\limits_{t=1}^{T} (\hat{\Fb}_{t}^\top \Fb_{0t})\Big)$. Then, under Assumptions 1-3,
		\[\dfrac{1}{p_1} \lVert \hat{\Rb} - \Rb_0 \hat{\Sbb}_1 \rVert_F^2 +\dfrac{1}{p_2} \lVert \hat{\Cb} - \Cb_0 \hat{\Sbb}_2 \rVert_F^2 +
		\dfrac{1}{T^2}\lVert \sum\limits_{t=1}^T (\hat{\Fb}_t - \hat{\Sbb}_1 \Fb_{0t} \hat{\Sbb}_2) \rVert_F^2 = O_p\left(\dfrac{1}{L}\right),
		\]
where  $L = \min\{p_1p_2, Tp_2, Tp_1\}$. In particular,
\[\dfrac{1}{p_1} \lVert \hat{\Rb} - \Rb_0 \hat{\Sbb}_1 \rVert^2 +\dfrac{1}{p_2} \lVert \hat{\Cb} - \Cb_0 \hat{\Sbb}_2 \rVert^2 +
		\dfrac{1}{T^2}\lVert \sum\limits_{t=1}^T (\hat{\Fb}_t - \hat{\Sbb}_1 \Fb_{0t} \hat{\Sbb}_2) \rVert^2 = O_p\left(\dfrac{1}{L}\right).
		\]
	\end{theorem}
	
In Theorem \ref{the1}, the sign matrices $\hat{\Sbb}_1,\hat{\Sbb}_2$ appear  due to the intrinsic sign indeterminacy of
factors and loadings, i.e., the factor structure remains the same when a factor and its
loading are both multiplied by -1.
Theorem \ref{the1} shows that the convergence rate of the least square estimator of $\theta$ is no slower than that of the $\alpha$-PCA estimator in \cite{fan2021}. In particular, when $p_1\ll Tp_2$, the theoretical minimizer converges at rate $O_p((Tp_1)^{-1}+(p_1p_2)^{-1})$, much faster than $O_p(p_1^{-1})$ of the $\alpha$-PCA estimator.  In terms of the summed errors of $\hat{\theta}$, i.e.,
$$
\frac{1}{p_1}\|\hat{\Rb}-\Rb_0\hat{\Sbb}_1\|_F^2+\frac{1}{p_2}\|\hat{\Cb}-\Cb_0\hat{\Sbb}_2\|_F^2+\frac{1}{T^2}\|\sum^T_{t=1}(\hat{\Fb}_t-\hat{\Sbb}_1\Fb_{0t}\hat{\Sbb}_2)\|_F^2,
$$
the least square estimate in Theorem \ref{the1} (also in Theorem \ref{the2}) has the same convergence rate as the PE estimate given in \cite{Yu2021Projected}: both are $O_p(L^{-1})$. However,  the convergence rate of the least square estimator for each individual component of $\theta$
derived in Theorem \ref{the1} is generally slower than that of the PE estimator in \cite{Yu2021Projected}, except that one of $((Tp_1)^{-1}, (Tp_2)^{-1}, (p_1p_2)^{-1})$ dominates the others. The reason is that $(\Rb_0, \Cb_0)$ are estimated jointly with $\{\Fb_{0t}, t=1,\ldots, T\}$ as the theoretical minimizers of the empirical square loss in the current work, while the two-step PE approach estimates $\Rb_0$ or $\Cb_0$ without knowing $\{\Fb_{0t}\}$ first. For example, the PE estimate of $\Rb_0$ converges at rate $O_p((Tp_2)^{-1}+(Tp_1)^{-2}+(p_1p_2)^{-2})$. When $(Tp_2)^{-1}$ dominates $(Tp_1)^{-1}$ and $(p_1p_2)^{-1}$, the PE estimate converges at the same rate as that derived in Theorem \ref{the1}. The reason is that the two-step estimate of $\Rb_0$ of the PE method depends on $\Cb_0$ in only one step and does not rely on $\{\Fb_{0t}, t=1,\ldots, T\}$, and thus can be estimated individually, while the least square minimizes the empirical loss function jointly in $\theta$.
\section{Robust Extension with the Huber Loss}

To account for the possible heavy-tailedness of the data distribution, we extend the above work by replacing the square loss by the Huber loss, and provide an efficient algorithm to solve the optimization problem. It turns out that the algorithm is simply a weighted version of Algorithm 1. We also provide an iterative algorithm to determine the numbers of the column/row factors.
	
\subsection{Algorithm}\label{sec:2.2}
	Standard statistical procedures that are based on the method of least squares often
behave poorly in the presence of heavy-tailed data. The observed data are often heavy-tailed in areas such as finance and macroeconomics. To deal with heavy-tailed data, a natural idea is to replace the least squares loss function with the Huber loss function, i.e., we consider the following optimization problem:

\begin{equation}\label{equ:Huberloss}
	\begin{array}{ccc}
		 \min_{\{\mathbf{R},\mathbf{C},\mathbf{F}_t\}}L_2(\mathbf{R},\mathbf{C},\mathbf{F}_t)=\dfrac{1}{T}\sum\limits_{t=1}^{T}H_{\tau}\Big(\sqrt{\Vert \mathbf{X}_t - \mathbf{R} \mathbf{F}_t \mathbf{C}^\top \Vert ^2 _F }\Big),\vspace{2ex} \\
		\text{s.t.}\dfrac{1}{p_1} \mathbf{R}^\top \mathbf{R} = \mathbf{I}_{k_1},\dfrac{1}{p_2} \mathbf{C}^\top \mathbf{C} = \mathbf{I}_{k_2}.
	\end{array}
\end{equation}

	where the Huber loss $H_{\tau}(x)$ is defined as
	$$ H_{\tau}(x)=
	\begin{cases}
	     x^2, & \text{$\vert x \vert \leq \tau$ }, \\
		2\tau \vert x \vert - \tau ^2, & \text{$\vert x \vert > \tau$}.
	\end{cases} $$

	By similar arguments as for the least squares case (see details in the supplementary material), we have
\begin{equation}\label{equ:Mrw}
\begin{array}{cc}
\mathbf{R} \mathbf{\Theta} =\Mb_{c}^{w} \mathbf{R}  \ \ \text{and} \ \ \mathbf{C} \mathbf{\Lambda}  =\mathbf{M}_r^w \mathbf{C}, \ \text{where}\vspace{2ex}\\
\Mb_{c}^{w}=\dfrac{1}{Tp_2}\sum\limits_{t=1}^{T} w_t \mathbf{X}_t \mathbf{C} \mathbf{C}^\top \mathbf{X}_t^\top,\ \ \mathbf{M}_r^w=\dfrac{1}{Tp_1}\sum\limits_{t=1}^{T} w_t \mathbf{X}_t^\top \mathbf{R} \mathbf{R}^\top \mathbf{X}_t,
\end{array}
\end{equation}
where $\mathbf{\Theta}$ and $\mathbf{\Lambda}$ are diagonal  Lagrangian multipliers  matrices and the weights are
	$$
	w_t=
	\left\{
\begin{array}{ccc}
		1, & \sqrt{\Vert \mathbf{X}_t - \mathbf{R} \mathbf{F}_t \mathbf{C}^T \Vert ^2 _F} \leq \tau, \vspace{2ex} \\
		\tau \dfrac{1}{\sqrt{\text{Tr}(\mathbf{X}_t^\top \mathbf{X}_t)- \dfrac{1}{p_1p_2} \text{Tr}(\mathbf{X}_t^\top \mathbf{R} \mathbf{R}^\top \mathbf{X}_t \mathbf{C} \mathbf{C}^\top)}}, & \sqrt{\Vert \mathbf{X}_t - \mathbf{R} \mathbf{F}_t \mathbf{C}^T \Vert ^2 _F} > \tau.
\end{array}
	\right.
	$$
	
	By (\ref{equ:Mrw}), we see that the $\Mb_c^w$ and $\Mb_r^w$ are weighted versions of the $\Mb_c$ and $\Mb_r$, respectively.
	We denote the first $k_1$ eigenvectors of $\Mb_c^w$ as $\{\br^w(1),\ldots,\br^w(k_1)\}$ and the corresponding eigenvalues as $\{\theta^w_1,\ldots,\theta^w_{k_1}\}$. Similarly, we denote the first $k_2$ eigenvectors of $\Mb_r^w$ as $\{\bc^w(1),\ldots,\bc^w(k_2)\}$ and the corresponding eigenvalues as $\{\lambda_1^w,\ldots,\lambda_{k_2}^w\}$.
    From (\ref{equ:Mrw}), we clearly see that  $\Rb^w=\sqrt{p_1}(\br^w(1),\ldots,\br^w(k_1))$, $\Cb^w=\sqrt{p_2}(\bc^w(1),\ldots,\bc^w(k_2))$, $\bTheta^w=\text{diag}(\theta_1^w,\ldots,\theta_{k_1}^w)$, $\bLambda^w=\text{diag}(\lambda_1^w,\ldots,\lambda_{k_2}^w)$ satisfy the KKT condition. Both $\Mb_c^w$ and $\Mb_r^w$  rely on the unknown row/column factor loadings (see the weights $w_t$).

    We propose an iterative procedure to get the estimators, which turns out to be slightly different from the iterative procedure in the last section, as we need to update $\Rb^w$ and $\Cb^w$ simultaneously to update the weights $w_t$. We summarized the algorithm in Algorithm \ref{alg:RMFA} and the initial estimators $\hat \Rb$ and $\hat\Cb$ can also be chosen as the estimators by $\alpha$-PCA. Once the factor loading matrices are estimated, the factor matrix $\Fb_t$ can be estimated by $ \widetilde \Fb_t^w=\mathbf{\widetilde{R}}^{w\top} \mathbf{X}_t \mathbf{\widetilde{C}}^w/(p_1p_2)$ and thus the common component matrix $\Sbb$ can be estimated by $\widetilde \Sbb^w=\mathbf{\widetilde{R}}^{w}\widetilde \Fb_t^w\mathbf{\widetilde{C}}^{w\top}$. As in each step, we need to update the weights, it is hard to derive the asymptotic theory for the estimators in Algorithm 2 and  we demonstrate its superiority by extensive numerical experiments. We develop the asymptotic theory for the theoretical minimizers in the following and leave the asymptotic theory of the estimators from Algorithm 2 to our future work.

\begin{algorithm}[!h]
		\caption{Robust Matrix Factor Analysis}\label{alg:RMFA}
		\hspace*{0.02in} {\bf Input:} Data matrices $\{\mathbf{X_t}\}_{t\leq T}$, the row factor number $k_1$, the column factor number $k_2$ \\
		\hspace*{0.02in} {\bf Output:} Factor loading matrices $\mathbf{\widetilde{R}}^w$ and $\mathbf{\widetilde{C}}^w$\\
		\begin{algorithmic}[1]
			\State Obtain the initial estimators $\mathbf{\hat{R}}^{(0)}$ and $\mathbf{\hat{C}}^{(0)}$ by $\alpha$-PCA;\\
			Compute the weights $\{w_t\}, t=1,\ldots,T$; \\
			Using $\{w_t\}$ and  $ \hat\Rb^{(0)}$ and $\hat\Cb^{(0)}$
			to compute $\Mb_c^w$ and its corresponding first $k_1$ eigenvectors $\{\br^w(1),\ldots,\br^w(k_1)\}$. Update $\hat \Rb^{(1)}$ as $\sqrt{p_1}(\br^w(1),\ldots,\br^w(k_1))$. \\
		Using $\{w_t\}$ and  $ \hat\Rb^{(0)}$ and $\hat\Cb^{(0)}$
			to compute $\Mb_r^w$ and its corresponding first $k_2$ eigenvectors $\{\bc^w(1),\ldots,\bc^w(k_2)\}$. Update $\hat \Cb^{(1)}$ as $\sqrt{p_2}(\bc^w(1),\ldots,\bc^w(k_2))$. \\
			Repeat steps 2-4 until convergence and output the estimators from the last step and denoted as $\tilde \Rb^w$ and $\tilde \Cb^w$.
		\end{algorithmic}
	\end{algorithm}

{In (\ref{equ:Yu}), when there exist outliers in observations $\{\Yb_t\}$, we naturally would consider a weighted  sample covariance matrix  to decrease the impact of outliers, and $\Mb_{c}^{w}$ would be an ideal choice. This turns out to be the estimators by minimizing the Huber loss function, which is exactly a weighted version of the projection technique.
The weighted projection technique not only  reduces the magnitudes of the
idiosyncratic error components and thus increases the signal-to-noise ratio, but also neglects the impact of outliers by putting  very small weights on them. As far as we know, this is the first time that the weighted projection technique is proposed in matrix/tensor-valued data analysis.}

In the following, we derive the convergence rates of the theoretical minimizers of the Huber loss function in (\ref{equ:Huberloss}), denoted by $(\hat{\Rb}_h, \hat{\Fb}_{th}, \hat{\Cb}_h)$. To this end, we introduce the following Assumption 3$^{\prime}$, parallel to the Assumption 3 in Section \ref{sec:4}.

\vspace{0.5em}

\textbf{Assumption 3$^{\prime}$: }  The distribution functions of $e_{ij,t}|\{\Fb_{0t}\}$ have a common support covering an open neighborhood of the origin, $T\log{T}/(p_1p_2)=o(1)$, $E(e_{ij,t}|\{\Fb_{0t}\})=0$ and $E((e_{ij,t})^{2+\epsilon}|\{\Fb_{0t}\})\leq C$ for some constant $C$ and any arbitrarily small number $\epsilon>0$.

   \vspace{0.5em}

Assumption 3$^{\prime}$ relaxes the  sub-Gaussian tail constraint of the idiosyncratic errors and only requires the boundedness of the $(2+\epsilon)$th moment, though with a mild scaling condition on $(p_1,p_2,T)$.
 Similar to Theorem \ref{the1}, we obtain the following theorem which establishes the same convergence rates either under Assumption 3 or Assumption 3$^{\prime}$.

   	\begin{theorem}\label{the2}
 Let	$\hat{\Sbb}_{1h} = \mbox{sgn}\Big(\dfrac{1}{T} \sum\limits_{t=1}^{T} (\hat{\Fb}_{th} \Fb_{0t}^\top)\Big)$ and $\hat{\Sbb}_{2h} = \mbox{sgn}\Big(\dfrac{1}{T} \sum\limits_{t=1}^{T} (\hat{\Fb}_{th}^\top \Fb_{0t})\Big)$. Further let $\tau= \tau^\prime\sqrt{p_1p_2}$ for some constant $\tau^\prime>0$, then either under Assumptions 1,2,3, or under Assumptions 1, 2, 3$^{\prime}$, we have
		\[\dfrac{1}{p_1} \lVert \hat{\Rb}_h - \Rb_0 \hat{\Sbb}_{1h} \rVert_F^2 + \dfrac{1}{p_2} \lVert \hat{\Cb}_h - \Cb_0 \hat{\Sbb}_{2h} \rVert_F^2 +
		\dfrac{1}{T^2}\lVert \sum\limits_{t=1}^T (\hat{\Fb}_{th} - \hat{\Sbb}_{1h} \Fb_{0t} \hat{\Sbb}_{2h}) \rVert_F^2 = O_p\left(\dfrac{1}{L}\right),
		\]
where  $L = \min\{p_1p_2, Tp_2, Tp_1\}$.
	\end{theorem}

The convergence rates in Theorem \ref{the2} are still correct for the spectral norm which is upper bounded by the Frobenius norm. The constant $\tau^{\prime}$ in the theorem serves as a tuning parameter that controls the fraction of data to be winsorized. As in \cite{huang2008robust}, we suggest to set $\tau^{\prime}$ so that a half of the data matrices $\{\Xb_t, t=1,\ldots,T\}$ are winsorized, which is justified by extensive simulation studies later.

\subsection{Determining the Pair of Factor Numbers}\label{sec:2.3}

It's well known that accurate estimation of the numbers of factors is of great importance to do matrix factor analysis \citep{Yu2021Projected}.
	We borrow the eigenvalue-ratio idea from \cite{ahn2013eigenvalue}. In detail,   $k_1$ and $k_2$ are estimated by
\begin{equation}\label{equ:eigenratio}
\hat k_1=\arg\max_{j\le k_{\max}}\frac{\lambda_j(\Mb_c^w)}{\lambda_{j+1}(\Mb_c^w)}, \ \ \hat k_2=\arg\max_{j\le k_{\max}}\frac{\lambda_j(\Mb_r^w)}{\lambda_{j+1}(\Mb_r^w)}
\end{equation}
where $k_{\max}$ is a predetermined value larger than $k_1$ and $k_2$.

	\begin{algorithm}[!h]
		\caption{Robust iterative algorithm to estimate the numbers of factors}\label{alg:fn}
		\hspace*{0.02in} {\bf Input:} Data $\mathbf{X_t}$, maximum number $k_{max}$, maximum iteration steps $m$\\
		\hspace*{0.02in} {\bf Output:} Numbers of row and column factors $\hat{k}_1$ and $\hat{k}_2$ \\
		\begin{algorithmic}[1]
			\State initialization: $\hat{k}^{(0)}_1 = k_{max},\hat{k}^{(0)}_2 = k_{max}$;\\
              estimate $\Rb$ and $\Cb$ by $\alpha$-PCA and denote the estimators as $\hat\Rb^{(0)}$ and $\hat\Cb^{(0)}$, respectively.\\
			for $t=1,2,\cdots,m$, given $\hat{k}^{(t-1)}_2$,  calculate $\Mb_c^{w(t)}$ using $\hat{\mathbf{R}}^{(t-1)}$ and $\hat{\mathbf{C}}^{(t-1)}$, then obtain $\hat{k}_1^{(t)}$ by (\ref{equ:eigenratio}) ;\\
			given $\hat{k}_1^{(t)}$, update $\hat{\mathbf{R}}^{(t)}$ by (\ref{equ:Mrw}), and calculate $\Mb_r^{w(t)}$ using $\hat{\mathbf{R}}^{(t)}$ and $\hat{\mathbf{C}}^{(t-1)}$, then obtain $\hat{k}_2^{(t)}$ by (\ref{equ:eigenratio});\\
            given $\hat{k}_2^{(t)}$, update $\hat{\mathbf{C}}^{(t)}$ by (\ref{equ:Mrw}).\\
			repeat steps 3-5 until $\hat{k}_1^{(t)}=\hat{k}_1^{(t-1)}$ and $\hat{k}_2^{(t)}=\hat{k}_2^{(t-1)}$ or reach the maximum iteration steps.
		\end{algorithmic}
	\end{algorithm}

If the common factors are sufficiently strong, the leading $k_1$ eigenvalues of $\Mb_c^w$ ($\Mb_r^w$) are well separated from the others, and the eigenvalue ratios in equation (\ref{equ:eigenratio}) will be asymptotically maximized exactly at $j=k_1$ ( $j=k_2$). To avoid vanishing denominators, we can add an asymptotically negligible term to the denominator of equation (\ref{equ:eigenratio}).  The remaining problem is that to calculate $\Mb_c^w$ or $\Mb_r^w$, both $\Rb$ and $\Cb$ must be predetermined, which further indicates that $k_1, k_2$ must be given in advance. However, both $k_1$ and $k_2$ are unknown empirically. To circumvent the problem, we propose an iterative algorithm  to determine the pair of   factor numbers, see Algorithm \ref{alg:fn}.
Thus our method is an \textbf{R}obust \textbf{it}erative \textbf{E}igenvalue-\textbf{R}atio (Rit-ER) based procedure.

	\section{Simulation Study}\label{sec:5}
	\subsection{Data Generation}
	 In this section, we introduce the data generating mechanism of the synthetic dataset in order to verify the performance of the proposed \textbf{R}obust-\textbf{M}atrix-\textbf{F}actor-\textbf{A}nalysis (RMFA) method in Algorithm 2.
	
	 We set $ k_1 = k_2 = 3 $, draw the entries of $\mathbf{R}_0$ and $\mathbf{C}_0$ independently from the uniform distribution $\mathcal{U}(-1,1)$, and let
	 \[
\begin{array}{cccc}
	 \text{Vec}(\mathbf{F}_{0t}) = \phi \times \text{Vec}(\mathbf{F}_{0,t-1}) + \sqrt{1-\phi^2} \times \mathbf{\epsilon}_t, \mathbf{\epsilon}_t \stackrel{i.i.d}{\sim}  \mathcal{N}(\mathbf{0}, \mathbf{I}_{k_1 \times k_2}),\vspace{1em}\\
	 \text{Vec}(\mathbf{E}_t) = \psi \times \text{Vec}(\mathbf{E}_{t-1}) + \sqrt{1-\psi^2} \times \text{Vec}(\mathbf{U}_t),
\end{array}
	 \]
	where $\phi$ and $\psi$ control the temporal and cross-sectional
correlations, and $\mathbf{U}_t$s are generated either from the matrix normal distribution or matrix $t$-distribution respectively. In detail, when $\mathbf{U}_t$ is from a matrix normal distribution $ \mathcal{MN} (\mathbf{0},\mathbf{U}_E,\mathbf{V}_E) $,
	then Vec($\mathbf{U}_t$) $\stackrel{i.i.d}{\sim}$ $\mathcal{N}$($\mathbf{0}$, $\mathbf{V}_E$ $\otimes$ $\mathbf{U}_E$).
	When $\mathbf{U}_t$ is from a matrix $t$-distribution $t_{p_1,p_2} (\nu,\mathbf{M},\mathbf{U}_E,\mathbf{V}_E) $, $\mathbf{U}_t$ has probability density function
	 \[f(\mathbf{U}_t;\nu,\mathbf{M},\mathbf{U}_E,\mathbf{V}_E)
	 =K \times \Big\vert \mathbf{I}_{p_1} \mathbf{U}_E^{-1} (\mathbf{U}_t -\mathbf{M}) \mathbf{V}_E^{-1} (\mathbf{U}_t -\mathbf{M})^\top \Big\vert^{-\dfrac{\nu+p_1+p_2-1}{2}},
	 \]
	 where $K$ is the regularization parameter. In our simulation study, we set $\Mb=0$ and let $\mathbf{U}_E$ and $\mathbf{V}_E$ be matrices with ones on the diagonal, and the off-diagonal entries are $ {1}/{p_1} $ and $ {1}/{p_2}$, respectively. For  matrix $t$-distribution, we resort to the \textbf{R} package ``MixMatrix" to generate random samples. The pair of factor numbers is given except in Section \ref{sec:fn}. We also investigate in the supplement the performance of various methods in reconstructing the common components with the factor numbers estimated by Algorithm 3, and we find no big difference with the case where $k_1$ and $k_2$ are known.
All the simulation results reported hereafter are based on 1000 replications.
	 As in \cite{huang2008robust}, the parameter $\tau$ is set as the median of $\left\{{\Vert \mathbf{X}_t - \hat\Rb \hat{\mathbf{F}}_t \hat\Cb^\top \Vert  _F },t=1\ldots,T\right\}$, where $\hat\Rb$ and $\hat\Cb$ are the initial estimators by the $\alpha$-PCA method with $\alpha=0$ and $\hat{\mathbf{F}}_t=\hat{\mathbf{R}}^\top \mathbf{X_t} \hat{\mathbf{C}}/(p_1p_2)$.

In the supplement, we also present the computation time of various methods in terms of estimating the loading spaces and the factor numbers. The computational burden of our RMFA method is sort of in the middle in all procedures given below.

	 \subsection{Estimation Error for the Loading Spaces}\label{sec:load}
	 In this section, we compare the RMFA method in Algorithm 2, the $\alpha$-PCA method  by \cite{fan2021}, the PE method by \cite{Yu2021Projected} in Algorithm 1, the Time series Outer-Product Unfolding Procedure (TOPUP), the Time series Inner-Product Unfolding Procedure (TIPUP)  both  by \cite{chen2022factor}, and the corresponding iterative version of TOPUP/TIPUP (denoted by iTOPUP/iTIPUP) by \cite{Han2020Tensor},   in terms of estimating the factor loading spaces. We consider the following two settings:

	\begin{itemize}
      \item  \textbf{Setting A}: $p_1=20,T=p_2 \in \{ 20,50,100,150,200 \}, \phi=\psi=0.1$.
      \item  \textbf{Setting B}: $p_2=20,T=p_1 \in \{ 20,50,100,150,200 \}, \phi=\psi=0.1$.
    \end{itemize}

\begin{table}[!h]
	 	\caption{Averaged estimation errors and standard errors of $\mathcal{D} (\hat{\mathbf{R}}, \mathbf{R})$ and $\mathcal{D} (\hat{\mathbf{C}}, \mathbf{C})$ for Settings A and B under Matrix Normal distribution and Matrix $t_3$ distribution over 1000 replications. ``RMFA": proposed robust matrix factor analysis method. ``$\alpha$-PCA": $\alpha$-PCA with $\alpha = 0$. ``PE": projection estimation method.  ``TOPUP": Time series Outer-Product Unfolding Procedure. ``TIPUP": Time series Inner-Product Unfolding Procedure. ``iTOPUP": iterative procedure based on the TOPUP. ``iTIPUP": iterative procedure based on the TIPUP.}
	 	 \label{tab:main1}\renewcommand{\arraystretch}{0.8} \centering
	 	\selectfont
	 	\begin{threeparttable}
	 		 \scalebox{0.7}{\begin{tabular*}{24.5cm}{ccccccccccccccccccccccccccccc}
	 				\toprule[2pt]
	 			    &\multirow{2}{*}{Evaluation}  &\multirow{2}{*}{$T$}
	 				&\multirow{2}{*}{$p_1$} &\multirow{2}{*}{$p_2$}
	 				&\multirow{2}{*}{RMFA}  &\multirow{2}{*}{$\alpha$-PCA}
	 				&\multirow{2}{*}{PE}   &\multirow{2}{*}{TOPUP}
	 				&\multirow{2}{*}{TIPUP} &\multirow{2}{*}{iTOPUP}
	 				 &\multirow{2}{*}{iTIPUP}   \cr
	 			     \\
	 				\midrule[1pt]
	 				 &&&& \multicolumn{7}{c}{\multirow{1}{*}{\textbf{Matrix Normal Distribution}}}\\[1ex]
	 				&$\mathcal{D} (\hat{\mathbf{R}}, \mathbf{R})$  &20 &20  &20  &$0.0915(0.0160)$  &$0.1117(0.0314) $ &$0.0919(0.0164) $  & $ 0.1488 ( 0.0398 ) $ & $ 0.4649 ( 0.1000 ) $ & $ 0.1727 ( 0.0390 ) $ & $ 0.4318 ( 0.1020 ) $\\
	 				&  &50  &  &50  &$0.0356(0.0052)$  &$0.0594(0.0233) $ &$0.0356(0.0052)$  & $ 0.0814 ( 0.0275 ) $ & $ 0.3845 ( 0.1219 ) $ & $ 0.0974 ( 0.0180 ) $ & $ 0.3442 ( 0.1115 ) $\\
	 				&  &100  &  &100  &$0.0176(0.0024)$  &$0.0467(0.0193) $ &$0.0176(0.0023) $ & $ 0.0605 ( 0.0207 ) $ & $ 0.3133 ( 0.1583 ) $ & $ 0.0658 ( 0.0122 ) $ & $ 0.2505 ( 0.1074 ) $\\
	 				&  &150  &  &150  &$0.0117(0.0016) $ &$0.0432(0.0196) $ &$0.0116(0.0016)$  & $ 0.0527 ( 0.0206 ) $ & $ 0.2367 ( 0.1632 ) $ & $ 0.0516 ( 0.0091 ) $ & $ 0.1858 ( 0.0995 ) $\\
	 				&  &200  &  &200  &$0.0088(0.0012)$  &$0.0421(0.0201) $ &$0.0088(0.0012) $ & $ 0.0489 ( 0.0209 ) $ & $ 0.2023 ( 0.1664 ) $ & $ 0.0435 ( 0.0080 ) $ & $ 0.1491 ( 0.0952 ) $\\
	 				&$\mathcal{D} (\hat{\mathbf{C}}, \mathbf{C})$
	 				&20 &20  &20  &$0.0930(0.0162) $ &$0.1140(0.0313) $ &$0.0935(0.0167) $ & $ 0.1501 ( 0.0401 ) $ & $ 0.4652 ( 0.1016 ) $ & $ 0.1731 ( 0.0375 ) $ & $ 0.4351 ( 0.1055 ) $\\
	 				&  &50  &  &50  &$0.0567(0.0060)$  &$0.0590(0.0066) $ &$0.0569(0.0061)$  & $ 0.0956 ( 0.0093 ) $ & $ 0.4131 ( 0.0891 ) $ & $ 0.1409 ( 0.0173 ) $ & $ 0.4073 ( 0.0925 ) $\\
	 				&  &100  &  &100  &$0.0398(0.0032) $ &$0.0404(0.0034) $ &$0.0399(0.0033) $ & $ 0.0823 ( 0.0055 ) $ & $ 0.3689 ( 0.0820 ) $ & $ 0.1301 ( 0.0129 ) $ & $ 0.3682 ( 0.0827 ) $\\
	 				&  &150  &  &150  &$0.0323(0.0025)$  &$0.0325(0.0025) $ &  $0.0324(0.0025)$ & $ 0.0772 ( 0.0045 ) $ & $ 0.3362 ( 0.0815 ) $ & $ 0.1240 ( 0.0114 ) $ & $ 0.3367 ( 0.0820 ) $\\
	 				&  &200  &  &200  &$0.0281(0.0021) $ &$0.0282(0.0021) $ & $0.0282(0.0022)$ & $ 0.0750 ( 0.0038 ) $ & $ 0.3060 ( 0.0789 ) $ & $ 0.1204 ( 0.0111 ) $ & $ 0.3071 ( 0.0792 ) $\\
	 				\cmidrule(lr){3-13}
	 				&$\mathcal{D} (\hat{\mathbf{R}}, \mathbf{R})$  &20 &20  &20  &$0.0915(0.0160) $ &$0.1117(0.0314)$  &$0.0919(0.0164)$ & $ 0.1488 ( 0.0398 ) $ & $ 0.4649 ( 0.1000 ) $ & $ 0.1727 ( 0.0390 ) $ & $ 0.4318 ( 0.1020 ) $\\
	 				&  &50  &50  &  &$0.0567(0.0061) $ &$0.0591(0.0069) $ &$0.0569(0.0063) $ & $ 0.0955 ( 0.0095 ) $ & $ 0.4162 ( 0.0937 ) $ & $ 0.1411 ( 0.0176 ) $ & $ 0.4114 ( 0.0948 ) $\\
	 				&  &100  &100  &  &$0.0399(0.0034)$  &$0.0405(0.0036) $ &$0.0400(0.0035)  $ & $ 0.0823 ( 0.0056 ) $ & $ 0.3698 ( 0.0821 ) $ & $ 0.1306 ( 0.0135 ) $ & $ 0.3698 ( 0.0847 ) $\\
	 				&  &150  &150  &  &$0.0323(0.0024) $ &$0.0325(0.0025) $ &$0.0324(0.0025) $ & $ 0.0771 ( 0.0045 ) $ & $ 0.3353 ( 0.0829 ) $ & $ 0.1239 ( 0.0116 ) $ & $ 0.3345 ( 0.0815 ) $\\
	 				&  &200  &200  &  &$0.0280(0.0020)$  &$0.0281(0.0020)$  &$0.0281(0.0021)$  & $ 0.0752 ( 0.0041 ) $ & $ 0.3068 ( 0.0781 ) $ & $ 0.1208 ( 0.0114 ) $ & $ 0.3081 ( 0.0796 ) $\\
	 				\cmidrule(lr){3-13}
	 				&$\mathcal{D} (\hat{\mathbf{C}}, \mathbf{C})$
	 				&20 &20  &20  &$0.0930(0.0162) $ &$0.1140(0.0313) $ &$0.0935(0.0167)$  & $ 0.1501 ( 0.0401 ) $ & $ 0.4652 ( 0.1016 ) $ & $ 0.1731 ( 0.0375 ) $ & $ 0.4351 ( 0.1055 ) $\\
	 				&  &50  &50  &  &$0.0356(0.0053) $ &$0.0589(0.0225)$ &$0.0356(0.0053)$ & $ 0.0802 ( 0.0241 ) $ & $ 0.3854 ( 0.1225 ) $ & $ 0.0970 ( 0.0178 ) $ & $ 0.3378 ( 0.1100 ) $ \\
	 				&  &100  &100  &  &$0.0177(0.0025)$  &$0.0479(0.0216) $ &0.0177(0.0025) & $ 0.0606 ( 0.0222 ) $ & $ 0.3148 ( 0.1588 ) $ & $ 0.0658 ( 0.0123 ) $ & $ 0.2551 ( 0.1105 ) $ \\
	 				&  &150  &150  &  &$0.0118(0.0017) $ &$0.0437(0.0246)$  &  $0.0118(0.0017)$ & $ 0.0537 ( 0.0277 ) $ & $ 0.2462 ( 0.1711 ) $ & $ 0.0516 ( 0.0099 ) $ & $ 0.1900 ( 0.1042 ) $\\
	 				&  &200  &200  &  &$0.0088(0.0012)$  &$0.0428(0.0204)$  &$ 0.0088(0.0012)$ & $ 0.0487 ( 0.0212 ) $ & $ 0.1891 ( 0.1555 ) $ & $ 0.0435 ( 0.0084 ) $ & $ 0.1443 ( 0.0933 ) $\\
                 	\midrule[1pt]
	 				 &&&& \multicolumn{7}{c}{\multirow{1}{*}{\textbf{Matrix $t_3$ Distribution}}}\\[1ex]
	 				&$\mathcal{D} (\hat{\mathbf{R}}, \mathbf{R})$  &20 &20  &20  &$0.1454(0.1416)$  &$0.3752(0.1713)$ &$0.2830(0.2004)$  & $ 0.4109 ( 0.1694 ) $ & $ 0.5368 ( 0.1118 ) $ & $ 0.3340 ( 0.2049 ) $ & $ 0.5004 ( 0.1282 ) $\\
	 				&  &50  &  &50  &$0.0433(0.0738)$  &$0.2959(0.1716)$ &$0.1437(0.1943)$  & $ 0.3285 ( 0.1768 ) $ & $ 0.4860 ( 0.1383 ) $ & $ 0.1956 ( 0.1941 ) $ & $ 0.4044 ( 0.1416 ) $\\
	 				&  &100  &  &100  &$0.0171(0.0315)$  &$0.2657(0.1620)$ &$0.0830(0.1630)$ & $ 0.2921 ( 0.1736 ) $ & $ 0.4451 ( 0.1649 ) $ & $ 0.1397 ( 0.1817 ) $ & $ 0.3307 ( 0.1554 ) $\\
	 				&  &150  &  &150  &$0.0111(0.0179)$  &$0.2488(0.1575) $ &$0.0684(0.1571) $ & $ 0.2728 ( 0.1700 ) $ & $ 0.4128 ( 0.1740 ) $ & $ 0.1137 ( 0.1672 ) $ & $ 0.2727 ( 0.1612 ) $\\
	 				&  &200  &  &200  &$0.0088(0.0226)$  &$0.2382(0.1513)$ &$0.0581(0.1475)$ & $ 0.2621 ( 0.1662 ) $ & $ 0.3754 ( 0.1799 ) $ & $ 0.1006 ( 0.1625 ) $ & $ 0.2300 ( 0.1655 ) $\\
	 				\cmidrule(lr){3-13}
	 				&$\mathcal{D} (\hat{\mathbf{C}}, \mathbf{C})$
	 				&20 &20  &20  &$0.1439(0.1407) $ &$0.3709(0.1733) $ &$0.2783(0.2008) $ & $ 0.4040 ( 0.1667 ) $ & $ 0.5327 ( 0.1158 ) $ & $ 0.3302 ( 0.2026 ) $ & $ 0.5002 ( 0.1278 ) $\\
	 				&  &50  &  &50  &$0.0613(0.0732)$  &$0.2154(0.1853) $ &$0.1740(0.1941)$ & $ 0.2696 ( 0.1911 ) $ & $ 0.4876 ( 0.1291 ) $ & $ 0.2348 ( 0.1880 ) $ & $ 0.4581 ( 0.1305 ) $\\
	 				&  &100  &  &100  &$0.0362(0.0347)$  &$0.1387(0.1651)$  &$0.1197(0.1720)$ & $ 0.2025 ( 0.1831 ) $ & $ 0.4464 ( 0.1368 ) $ & $ 0.1989 ( 0.1755 ) $ & $ 0.4161 ( 0.1293 ) $\\
	 				&  &150  &  &150  &$0.0288(0.0166)$  &$0.1125(0.1588)$  & $ 0.0986(0.1621)$ & $ 0.1749 ( 0.1719 ) $ & $ 0.4056 ( 0.1374 ) $ & $ 0.1812 ( 0.1588 ) $ & $ 0.3867 ( 0.1279 ) $\\
	 				&  &200  &  &200  &$0.0253(0.0242)$  &$0.0992(0.1535) $ &$ 0.0892(0.1581)$ & $ 0.1608 ( 0.1657 ) $ & $ 0.3713 ( 0.1413 ) $ & $ 0.1726 ( 0.1538 ) $ & $ 0.3538 ( 0.1308 ) $\\
	 				\cmidrule(lr){3-13}
	 				&$\mathcal{D} (\hat{\mathbf{R}}, \mathbf{R})$  &20 &20  &20  &$0.1454(0.1416)$  &$0.3752 ( 0.1713 )$ &$0.2830 ( 0.2004 )$  & $ 0.4109 ( 0.1694 ) $ & $ 0.5368 ( 0.1118 ) $ & $ 0.3340 ( 0.2049 ) $ & $ 0.5004 ( 0.1282 ) $\\
	 				&  &50  &50  &  &$0.0598(0.0660)$  &$0.2143 ( 0.1837 )$ &$0.1734 ( 0.1925 )$  & $ 0.2618 ( 0.1896 ) $ & $ 0.4902 ( 0.1254 ) $ & $ 0.2324 ( 0.1876 ) $ & $ 0.4577 ( 0.1266 ) $\\
	 				&  &100  &100 &  &$0.0359(0.0237)$  &$0.1362 ( 0.1578 )$ &$0.1142 ( 0.1621 )$ & $ 0.1995 ( 0.1760 ) $ & $ 0.4467 ( 0.1315 ) $ & $ 0.2014 ( 0.1695 ) $ & $ 0.4225 ( 0.1266 ) $ \\
	 				&  &150  &150  &  &$0.0293(0.0231)$  &$0.1120 ( 0.1539 ) $ &$0.0972 ( 0.1572 ) $ & $ 0.1784 ( 0.1756 ) $ & $ 0.4109 ( 0.1430 ) $ & $ 0.1887 ( 0.1666 ) $ & $ 0.3858 ( 0.1320 ) $\\
	 				&  &200  &200  &  &$0.0247(0.0176)$  &$0.0923 ( 0.1447 )$ &$0.0815 ( 0.1473 )$ & $ 0.1625 ( 0.1724 ) $ & $ 0.3697 ( 0.1418 ) $ & $ 0.1760 ( 0.1570 ) $ & $ 0.3524 ( 0.1302 ) $ \\
	 				\cmidrule(lr){3-13}
	 				&$\mathcal{D} (\hat{\mathbf{C}}, \mathbf{C})$
	 				&20 &20  &20  &$0.1439(0.1407) $ &$0.3709 ( 0.1733 ) $ &$0.2783 ( 0.2008 ) $ & $ 0.4040 ( 0.1667 ) $ & $ 0.5327 ( 0.1158 ) $ & $ 0.3302 ( 0.2026 ) $ & $ 0.5002 ( 0.1278 ) $\\
	 				&  &50  &50  &  &$0.0418(0.0669)$  &$0.2992 ( 0.1742 ) $ &$0.1442 ( 0.1922 )$ & $ 0.3258 ( 0.1773 ) $ & $ 0.4872 ( 0.1329 ) $ & $ 0.1933 ( 0.1913 ) $ & $ 0.4072 ( 0.1388 ) $ \\
	 				&  &100  &50  &  &$0.0168(0.0231)$  &$0.2644 ( 0.1584 )$  &$0.0802 ( 0.1578 )$ & $ 0.2908 ( 0.1678 ) $ & $ 0.4473 ( 0.1608 ) $ & $ 0.1424 ( 0.1774 ) $ & $ 0.3307 ( 0.1530 ) $ \\
	 				&  &150  &150  &  &$0.0115(0.0252)$  &$0.2572 ( 0.1571 )$  & $ 0.0652 ( 0.1518 )$ & $ 0.2803 ( 0.1687 ) $ & $ 0.4068 ( 0.1752 ) $ & $ 0.1211 ( 0.1768 ) $ & $ 0.2704 ( 0.1635 ) $\\
	 				&  &200  &200  &  &$0.0082(0.0153)$  &$0.2459 ( 0.1559 ) $ &$ 0.0521 ( 0.1408 )$ & $ 0.2685 ( 0.1703 ) $ & $ 0.3865 ( 0.1813 ) $ & $ 0.1055 ( 0.1674 ) $ & $ 0.2254 ( 0.1618 ) $\\
	 				\bottomrule[2pt]
 				\end{tabular*}}
 			\end{threeparttable}
     \end{table}

We  first introduce a metric between two factor spaces as the factor loading matrices $\Rb$ and $\Cb$ are not identifiable. For two column-wise orthogonal matrices $(\bQ_1)_{p\times q_1}$ and $(\bQ_2)_{p\times q_2}$, we define
\[
\mathcal{D}(\bQ_1,\bQ_2)=\bigg(1-\frac{1}{\max{(q_1,q_2)}}\mbox{Tr}\Big(\bQ_1\bQ_1^{\top}\bQ_2\bQ_2^{\top}\Big)\bigg)^{1/2}.
\]
By the definition of $\mathcal{D}(\bQ_1,\bQ_2)$, we can easily see that its value  lies in the interval $[0,1]$, which measures the distance between the column spaces spanned by  $\bQ_1$ and $\bQ_2$. The column spaces spanned by $\bQ_1$ and $\bQ_2$  are the same when $\mathcal{D}(\bQ_1,\bQ_2)=0$, and are orthogonal when $\mathcal{D}(\bQ_1,\bQ_2)=1$.  The Gram-Schmidt orthonormal transformation can be used when $\bQ_1$ and $\bQ_2$ are not column-orthogonal matrices.

Table \ref{tab:main1} shows the averaged estimation errors with standard errors in parentheses under Settings A and B for matrix normal distribution and matrix-variate $t_3$ distribution. \cite{Yu2021Projected}'s simulation study showed that for the $\alpha$-PCA method, the performances for $\alpha\in\{-1,0,1\}$ are comparable, thus we only report the simulation results for the $\alpha$-PCA with $\alpha=0$. The TOPUP/TIPUP by \cite{chen2022factor} and iTOPUP/iTIPUP by \cite{Han2020Tensor} all performed worse than the other three methods, as these methods all rely on the strong persistency of the factors. All methods benefit from large dimensions, and when $p_1$ is small, RMFA and PE methods always show advantage over  $\alpha$-PCA in terms of estimating $\Rb$. When $p_2$ is small, RMFA and PE methods always show advantage over  $\alpha$-PCA in terms of estimating $\Cb$, which is consistent with the findings by \cite{Yu2021Projected}. What we want to emphasize is that the RMFA and PE method perform comparably under the normal idiosyncratic error case, which is also clearly seen from Figure \ref{fig:1} in the Introduction section. When the idiosyncratic errors are from heavy-tailed matrix $t_3$ distribution, the RMFA method is superior over the $\alpha$-PCA and PE methods in all scenarios. The estimation errors by the PE and $\alpha$-PCA  methods are at least twice of those by the RMFA method, which indicates that the weights of the sample covariance matrix of the projected data involved in RMFA  method play an important role when outliers exist. The simulation results for matrix $t_4$ distribution are put in Table 1 in the supplement and similar conclusions are drawn as for matrix $t_3$ distribution. As a result, the RMFA can be used as a safe replacement of the $\alpha$-PCA and PE methods.

	 \subsection{Estimation Error for Common Components}\label{sec:comcom}
 In this section, we compare the performances of the RMFA method with those of the $\alpha$-PCA  and PE methods in terms of estimating the common component matrices.
We evaluate the performance of different methods by  the Mean Squared Error, i.e.,
\[
\text{MSE}=\frac{1}{Tp_1p_2}\sum_{t=1}^T\|\hat\Sbb_t-\Sbb_t\|_F^2,
\]
where $\hat\Sbb_t$ refers to an arbitrary estimate and  $\Sbb_t$ is the true common component matrix at time point $t$.

Table \ref{tab:main2} shows the averaged MSEs with standard errors in parentheses under Settings A and B for matrix normal distribution and matrix-variate $t_3$ and $t_4$ distributions.
From Table \ref{tab:main2}, we see that the RMFA and PE methods perform comparably under the normal case, and both perform better than the $\alpha$-PCA method. This attributes to the projection technique of both the RMFA and PE methods. In contrast, under the heavy-tailed $t_3$ and $t_4$ cases, the RMFA  performs much better than both PE and $\alpha$-PCA methods. The TOPUP/TIPUP  and iTOPUP/iTIPUP all performed worse than the RMFA, PE and $\alpha$-PCA  methods.

\begin{table}[!h]
	 	\caption{Mean squared error and
its standard deviation of the common component under Settings A and B over 1000
replications. ``RMFA": proposed robust matrix factor analysis method. ``$\alpha$-PCA": $\alpha$-PCA with $\alpha = 0$. ``PE": projection estimation method. ``TOPUP": Time series Outer-Product Unfolding Procedure. ``TIPUP": Time series Inner-Product Unfolding Procedure. ``iTOPUP": iterative procedure based on the TOPUP. ``iTIPUP": iterative procedure based on the TIPUP.}
	 	 \label{tab:main2}\renewcommand{\arraystretch}{0.8} \centering
	 	\selectfont
	 	
	 	\begin{threeparttable}
	 		 \scalebox{0.75}{\begin{tabular*}{22.5cm}{ccccccccccccccccccccccccccccc}
	 				\toprule[2pt]
	 				&\multirow{2}{*}{Distribution}  &\multirow{2}{*}{$T$}
	 				&\multirow{2}{*}{RMFA}  &\multirow{2}{*}{$\alpha$-PCA}
	 				&\multirow{2}{*}{PE}    &\multirow{2}{*}{TOPUP}
	 				&\multirow{2}{*}{TIPUP} &\multirow{2}{*}{iTOPUP}
	 				&\multirow{2}{*}{iTIPUP} \cr
	 				\cmidrule(lr){8-11} \\
	 				\midrule[1pt]
	 				  \multicolumn{10}{c}{\multirow{1}{*}{Setting A: $p_1=20$, $T=p_2 $}}\\[1ex]
	 				\cmidrule(lr){2-13}
	 				&normal  &20   &$0.0136(0.0021)$  &$0.0188(0.0042)$ &$0.0137 ( 0.0022 )$ &$ 0.0333 ( 0.0085 )$ &$ 0.3346 ( 0.1026 )$ &$ 0.0475 ( 0.0123 )$ &$ 0.2931 ( 0.1047 )$ \\
	 				&  &50  &$0.0040( 0.0004 )$  &$0.0058 ( 0.0014 )$ &$0.0040( 0.0004 )$  &$ 0.0135 ( 0.0027 )$ &$ 0.2774 ( 0.1049 )$ &$ 0.0261 ( 0.0049 )$ &$ 0.2508 ( 0.0977 )$\\
	 				&  &100    &$0.0018 ( 0.0001 )$  &$0.0032 (0.0010)$ &$0.0018 ( 0.0002 )$  &$ 0.0094 ( 0.0018 )$ &$ 0.2250 ( 0.1141 )$ &$ 0.0199 ( 0.0033 )$ &$ 0.1887 ( 0.0814 )$\\
	 				&  &150   &$0.0011 ( 0.0001 )$  &$0.0025 ( 0.0009) $ &$0.0011 ( 0.0001 ) $ &$ 0.0080 ( 0.0015 )$ &$ 0.1774 ( 0.1146 )$ &$ 0.0172 ( 0.0027 )$ &$ 0.1491 ( 0.0764 )$ \\
	 				&  &200   &$0.0008 ( 0.0001 )$  &$0.0022 ( 0.0009 )$ &$0.0008 ( 0.0001 )$ &$ 0.0074 ( 0.0014 )$ &$ 0.1472 ( 0.1096 )$ &$ 0.0157 ( 0.0025 )$ &$ 0.1200 ( 0.0675 )$ \\
	 				\cmidrule(lr){2-13}
	 				&$t_4$  &20   &$ 0.0102 ( 0.0250 )$ &$ 0.0519 ( 0.0631 )$ &$ 0.0270 ( 0.0589 )$ &$ 0.0835 ( 0.0835 )$ &$ 0.3104 ( 0.1251 )$ &$ 0.0582 ( 0.0930 )$ &$ 0.2664 ( 0.1209 )$ \\
	 				&  &50   &$ 0.0019 ( 0.0004 )$ &$ 0.0219 ( 0.0499 )$ &$ 0.0101 ( 0.0444 )$ &$ 0.0363 ( 0.0629 )$ &$ 0.2619 ( 0.1279 )$ &$ 0.0275 ( 0.0650 )$ &$ 0.2162 ( 0.1121 )$ \\
	 				&  &100   &$ 0.0008 ( 0.0001 )$ &$ 0.0098 ( 0.0264 )$ &$ 0.0028 ( 0.0229 )$ &$ 0.0201 ( 0.0420 )$ &$ 0.2145 ( 0.1349 )$ &$ 0.0163 ( 0.0429 )$ &$ 0.1603 ( 0.1014 )$\\
	 				&  &150   &$ 0.0005 ( 0.0001 )$ &$ 0.0083 ( 0.0273 )$ &$ 0.0023 ( 0.0257 )$ &$ 0.0161 ( 0.0373 )$ &$ 0.1667 ( 0.1275 )$ &$ 0.0134 ( 0.0380 )$ &$ 0.1195 ( 0.0844 )$\\
	 				&  &200   &$ 0.0004 ( 0.0001 )$ &$ 0.0070 ( 0.0228 )$ &$ 0.0018 ( 0.0216 )$ &$ 0.0151 ( 0.0364 )$ &$ 0.1333 ( 0.1212 )$ &$ 0.0127 ( 0.0413 )$ &$ 0.0936 ( 0.0781 )$\\
	 				\cmidrule(lr){2-13}
	 				&$t_3$
	 				&20   &$0.0408(0.0743)$ &$0.1762(0.1242) $ &$0.1228(0.1354) $ &$ 0.2197 ( 0.1426 )$ &$ 0.4219 ( 0.1344 )$  &$ 0.1815 ( 0.1849 )$ &$ 0.3736 ( 0.1490 )$\\
	 				&  &50    &$0.0100(0.0500)$  &$0.1181(0.1371) $  & $0.0745(0.1403)$ &$ 0.1591 ( 0.1654 )$ &$ 0.3909 ( 0.1520 )$ &$ 0.1137 ( 0.1858 )$ &$ 0.3204 ( 0.1520 )$\\
	 				&  &100    &$0.0030(0.0389)$  &$0.0890(0.1232)$  & $0.0499(0.1269)$ &$ 0.1268 ( 0.1607 )$ &$ 0.3526 ( 0.1675 )$ &$ 0.0902 ( 0.1836 )$ &$ 0.2672 ( 0.1580 )$\\
	 				&  &150    &$0.0013(0.0124)$  &$0.0766(0.1152)$  &       $ 0.0420(0.1177)$ &$ 0.1081 ( 0.1474 )$ &$ 0.3052 ( 0.1655 )$ &$ 0.0729 ( 0.1604 )$ &$ 0.2221 ( 0.1520 )$\\
	 				&  &200    &$0.0014(0.0179)$  &$0.0708(0.1127) $ &       $ 0.0381(0.1136)$ &$ 0.1002 ( 0.1446 )$ &$ 0.2658 ( 0.1649 )$ &$ 0.0672 ( 0.1574 )$ &$ 0.1880 ( 0.1524 )$\\
	 				\midrule[1pt]
	 			    \multicolumn{10}{c}{\multirow{1}{*}{Setting B: $ p_2=20$, $T=p_1 $}}\\[1ex]
	 				\cmidrule(lr){2-13}
	 				&normal  &20   &$0.0136(0.0021)$  &$0.0188(0.0042)$ &$0.0137(0.0022)$ &$ 0.0333 ( 0.0085 )$ &$ 0.3346 ( 0.1026 )$  &$ 0.0475 ( 0.0123 )$ &$ 0.2931 ( 0.1047 )$ \\
	 				&  &50  &$0.0040(0.0004)$  &$0.0058(0.0014)$ &$0.0040(0.0004)$ &$ 0.0133 ( 0.0026 )$ &$ 0.2807 ( 0.1075 )$  &$ 0.0260 ( 0.0049 )$ &$ 0.2490 ( 0.0955 )$ \\
	 				&  &100    &$0.0018(0.0002)$  &$0.0033(0.0011)$ &$0.0018(0.0002)$ &$ 0.0093 ( 0.0018 )$ &$ 0.2247 ( 0.1144 )$ &$ 0.0198 ( 0.0033 )$ &$ 0.1906 ( 0.0848 )$ \\
	 				&  &150   &$0.0011(0.0001)$  &$0.0025(0.0013) $ &$0.0011(0.0001) $ &$ 0.0081 ( 0.0020 )$ &$ 0.1826 ( 0.1195 )$ &$ 0.0172 ( 0.0027 )$ &$ 0.1497 ( 0.0784 )$ \\
	 				&  &200   &$0.0008(0.0001)$  &$0.0022(0.0009)$ &$0.0008(0.0001)$ &$ 0.0074 ( 0.0015 )$  &$ 0.1413 ( 0.1052 )$ &$ 0.0158 ( 0.0027 )$ &$ 0.1200 ( 0.0698 )$ \\
	 				\cmidrule(lr){2-13}
	 				&$t_4$  &20   &$ 0.0102 ( 0.0250 )$ &$ 0.0519 ( 0.0631 )$ &$ 0.0270 ( 0.0589 )$ &$ 0.0835 ( 0.0835 )$ &$ 0.3104 ( 0.1251 )$ &$ 0.0582 ( 0.0930 )$ &$ 0.2664 ( 0.1209 )$ \\
	 				&  &50  &$ 0.0020 ( 0.0005 )$ &$ 0.0187 ( 0.0394 )$ &$ 0.0077 ( 0.0344 )$ &$ 0.0342 ( 0.0546 )$ &$ 0.2561 ( 0.1251 )$ &$ 0.0260 ( 0.0589 )$ &$ 0.2155 ( 0.1140 )$  \\
	 				&  &100  &$ 0.0008 ( 0.0001 )$ &$ 0.0112 ( 0.0328 )$ &$ 0.0041 ( 0.0300 )$ &$ 0.0208 ( 0.0428 )$ &$ 0.2107 ( 0.1348 )$ &$ 0.0160 ( 0.0403 )$ &$ 0.1616 ( 0.1005 )$  \\
	 				&  &150  &$ 0.0005 ( 0.0001 )$ &$ 0.0092 ( 0.0299 )$ &$ 0.0029 ( 0.0269 )$ &$ 0.0180 ( 0.0453 )$ &$ 0.1736 ( 0.1353 )$ &$ 0.0161 ( 0.0482 )$ &$ 0.1256 ( 0.0944 )$ \\
	 				&  &200  &$ 0.0004 ( 0.0001 )$ &$ 0.0072 ( 0.0241 )$ &$ 0.0017 ( 0.0216 )$ &$ 0.0149 ( 0.0377 )$ &$ 0.1401 ( 0.1203 )$ &$ 0.0129 ( 0.0377 )$ &$ 0.0968 ( 0.0751 )$ \\
	 				\cmidrule(lr){2-13}
	 				&$t_3$
	 				&20   &$0.0408(0.0743) $ &$0.1762( 0.1242 ) $ &$0.1228(0.1354) $ &$ 0.2197 ( 0.1426 )$ &$ 0.4219 ( 0.1344 )$ &$ 0.1815 ( 0.1849 )$ &$ 0.3736 ( 0.1490 )$\\
	 				&  &50    &$0.0087(0.0418)$  &$0.1160(0.1291) $ &$0.0723(0.1341)$  &$ 0.1520 ( 0.1564 )$ &$ 0.3916 ( 0.1440 )$ &$ 0.1097 ( 0.1774 )$ &$ 0.3198 ( 0.1501 )$\\
	 				&  &100    &$0.0021(0.0163)$  &$0.0843(0.1121)$  &$0.0447(0.1136)$ &$ 0.1195 ( 0.1428 )$ &$ 0.3457 ( 0.1557 )$ &$ 0.0875 ( 0.1647 )$ &$ 0.2656 ( 0.1488 )$ \\
	 				&  &150    &$0.0017(0.0178)$  &$0.0776(0.1137)$  & $0.0392(0.1139)$ &$ 0.1117 ( 0.1482 )$ &$ 0.3015 ( 0.1655 )$ &$ 0.0778 ( 0.1630 )$ &$ 0.2185 ( 0.1531 )$\\
	 				&  &200    &$0.0010(0.0120)$  &$0.0707(0.1129) $ &$0.0349(0.1149)$ &$ 0.1043 ( 0.1513 )$ &$ 0.2653 ( 0.1661 )$  &$ 0.0710 ( 0.1609 )$ &$ 0.1828 ( 0.1531 )$\\
	 				\bottomrule[2pt]
	 		\end{tabular*}}
	 	\end{threeparttable}
	 	
	 \end{table}

	 \subsection{Estimating the Numbers of Factors}\label{sec:fn}
	In this
subsection, we compare the empirical performances of the proposed Rit-ER method with those of the $\alpha$-PCA based ER method ($\alpha$-PCA-ER) by \cite{fan2021},  the IterER method by
\cite{Yu2021Projected}, the  information-criterion methods iTOP-IC/iTIP-IC based on iTOPUP/iTIPUP by \cite{han2020rank},  the eigenvalue-ratio methods iTOP-ER/iTIP-ER based on iTOPUP/iTIPUP by \cite{han2020rank} and the {\textbf T}otal mode-$k$ {\textbf {Cor}}relation {\textbf {Th}}resholding method (TCorTh) by \cite{lam2021rank} in terms of estimating the numbers of factors.

 \begin{table}[htb]
	 	\caption{The frequencies of exact estimation and underestimation of the numbers of factors under Settings A and B over 1000 replications. ``Rit-ER": the proposed robust iterative eigenvalue-ration based method. ``$\alpha$-PCA-ER": $\alpha$-PCA based eigenvalue-ratio method with $\alpha = 0$. ``IterER": iterative eigenvalue-ration based method. ``iTOP-IC"/``iTIP-IC":  information-criterion method based on iTOPUP/ iTIPUP.  ``iTOP-ER"/``iTIP-ER":  eigenvalue-ratio method based on iTOPUP/iTIPUP.  ``TCorTh": Total mode-$k$ Correlation Thresholding method.}
	 	 \label{tab:main3}\renewcommand{\arraystretch}{0.9} \centering
	 	\selectfont
	 	
	 	\begin{threeparttable}
	 		 \scalebox{0.75}{\begin{tabular*}{22.5cm}{ccccccccccccccccccccccccccccc}
	 				\toprule[2pt]
	 				&\multirow{2}{*}{Distribution}  &\multirow{2}{*}{$T$}
	 				&\multirow{2}{*}{Rit-ER}  &\multirow{2}{*}{$\alpha$-PCA-ER}
	 				&\multirow{2}{*}{IterER}  	 &\multirow{2}{*}{iTOP-IC}  &\multirow{2}{*}{iTIP-IC}
	 				&\multirow{2}{*}{iTOP-ER}  &\multirow{2}{*}{iTIP-ER} &\multirow{2}{*}{TCorTh}  \cr
	 				\cmidrule(lr){2-18} \\
	 				\midrule[1pt]
	 					  \multicolumn{11}{c}{\multirow{1}{*}{Setting A: $p_1=20$, $T=p_2 $}}\\[1ex]
	 				\cmidrule(lr){2-18}
	 				&normal  &20   &$0.992(0.008)$  &$0.647(0.353)$ &$0.992(0.008)$ &$0.000(1.000)$ &$0.000(1.000)$ &$0.758 (0.242)$ &$0.093 (0.907)$ &$0.672 (0.328)$\\
	 				&  &50  &$1.000(0.000)$  &$0.891(0.109)$ &$1.000(0.000)$ &$0.000(1.000)$ &$0.000(1.000)$ &$0.973 (0.027)$ &$0.196 (0.804)$ &$ 0.982  (0.018)$\\
	 				&  &100    &$1.000(0.000)$  &$0.880(0.120)$ &$1.000(0.000)$ &$0.000(1.000)$ &$0.000(1.000)$ &$0.994 (0.006)$ &$0.371 (0.629)$ &$ 0.997  (0.003)$\\
	 				&  &150   &$1.000(0.000)$  &$0.912(0.088) $ &$1.000(0.000) $ &$0.000(1.000)$ &$0.000(1.000)$ &$0.998 (0.002)$ &$0.525 (0.475)$ &$ 0.999  (0.001)$\\
	 				&  &200   &$1.000(0.000)$  &$0.901(0.099)$ &$1.000(0.000)$  &$0.000(1.000)$ &$0.000(1.000)$ &$0.996 (0.004)$ &$0.648 (0.352)$ &$ 0.999  (0.001)$\\
	 				\cmidrule(lr){2-18}
	 		
	 				&$t_4$  &20   &$0.774( 0.226)$  &$0.310(0.690)$ &$0.720(0.280)$ &$0.000(1.000)$ &$0.000(1.000)$ &$0.565 (0.435)$ &$0.112 (0.888)$ &$0.765 (0.235)$ \\
	 				&  &50   &$0.937( 0.063)$  &$0.622(0.378)$ &$0.864(0.136)$ &$0.000(1.000)$ &$0.000(1.000)$ &$0.795 (0.205)$ &$0.224 (0.776)$ &$ 0.414 (0.586)$ \\
	 				&  &100   &$0.986(0.014)$  &$0.665(0.335)$ &$0.933(0.067)$ &$0.000(1.000)$ &$0.000(1.000)$ &$0.888 (0.112)$ &$0.402 (0.598)$ &$ 0.148 (0.852)$ \\
	 				&  &150   &$0.992 ( 0.008)$  &$0.655(0.345) $ &$0.955(0.045) $ &$0.000(1.000)$ &$0.000(1.000)$ &$0.879 (0.121)$ &$0.581 (0.419)$ &$ 0.065 (0.935)$\\
	 				&  &200   &$0.991 (0.009)$  &$0.651(0.349)$ &$0.953(0.047)$  &$0.000(1.000)$ &$0.000(1.000)$ &$0.879 (0.121)$ &$0.685 (0.315)$ &$ 0.034 (0.966)$\\
	            \cmidrule(lr){2-18}
	 			&$t_3$
	 				&20   &$0.360(0.640) $ &$0.097( 0.903) $ &$0.307( 0.693) $ &$0.001(0.999)$ &$0.000(1.000)$ &$0.263 (0.737)$ &$0.050 (0.950)$ &$0.479 (0.521)$\\
	 				&  &50    &$0.704 ( 0.296)$  &$0.249( 0.751) $ &$0.579( 0.421)$  &$0.001(0.999)$ &$0.000(1.000)$ &$0.496 (0.504)$ &$0.155 (0.845)$ &$0.022 (0.978)$\\
	 				&  &100    &$0.810( 0.190)$  &$0.264( 0.736)$  &$0.643( 0.357)$ &$0.000(1.000)$ &$0.000(1.000)$ &$0.555 (0.445)$ &$0.266 (0.734)$ &$0.000 (1.000)$\\
	 				&  &150    &$0.855( 0.145)$  &$0.276( 0.724)$  & $ 0.691(0.309)$ &$0.000(1.000)$ &$0.000(1.000)$ &$0.611 (0.389)$ &$0.391 (0.609)$ &$0.000 (1.000)$\\
	 				&  &200    &$0.873( 0.127)$  &$0.260( 0.740) $ &$0.695(0.305)$ &$0.000(1.000)$ &$0.000(1.000)$ &$0.609 (0.391)$ &$0.470 (0.530)$ &$0.000 (1.000)$\\
	 				\midrule[1pt]
	 			 \multicolumn{11}{c}{\multirow{1}{*}{Setting B: $ p_2=20$, $T=p_1 $}}\\[1ex]
	 				\cmidrule(lr){2-18}
	 				&normal  &20   &$0.992(0.008)$  &$0.647(0.353)$ &$0.992(0.008)$ &$0.000(1.000)$ &$0.000(1.000)$ &$0.758 (0.242)$ &$0.093 (0.907)$ &$0.672 (0.328)$\\
	 				&  &50  &$1.000(0.000)$  &$0.895(0.105)$ &$1.000(0.000)$ &$0.000(1.000)$ &$0.000(1.000)$ &$0.988 (0.012)$ &$0.187 (0.813)$ &$0.978 (0.022)$\\
	 				&  &100    &$1.000(0.000)$  &$0.907(0.093)$ &$1.000(0.000)$ &$0.000(1.000)$ &$0.000(1.000)$ &$1.000 (0.000)$ &$0.343 (0.657)$ &$0.995 (0.005)$\\
	 				&  &150   &$1.000(0.000)$  &$0.909(0.091) $ &$1.000(0.000) $ &$0.000(1.000)$ &$0.000(1.000)$ &$0.998 (0.002)$ &$0.534 (0.466)$ &$0.999 (0.001)$\\
	 				&  &200   &$1.000(0.000)$  &$0.903(0.097)$ &$1.000(0.000)$  &$0.000(1.000)$ &$0.000(1.000)$ &$0.999 (0.001)$ &$0.646 (0.354)$ &$0.997 (0.003)$\\
	 				\cmidrule(lr){2-18}
	 				&$t_4$  &20   &$0.774(0.226)$  &$0.310(0.690)$ &$0.720(0.280)$  &$0.000(1.000)$ &$0.000(1.000)$ &$0.565 (0.435)$ &$0.112 (0.888)$ &$0.765 (0.235)$\\
	 				&  &50   &$0.926 (0.074)$  &$0.643(0.357)$ &$0.871(0.129)$ &$0.000(1.000)$ &$0.000(1.000)$ &$0.826 (0.174)$ &$0.248 (0.752)$ &$0.403 (0.597)$\\
	 				&  &100   &$0.950(0.050)$  &$0.672(0.328)$ &$0.913(0.087)$ &$0.000(1.000)$ &$0.000(1.000)$ &$0.884 (0.116)$ &$0.416 (0.584)$ &$0.171 (0.829)$\\
	 				&  &150   &$0.963(0.037)$  &$0.652(0.348) $ &$0.928(0.072) $ &$0.000(1.000)$ &$0.000(1.000)$ &$0.908 (0.092)$ &$0.532 (0.468)$ &$0.049 (0.951)$\\
	 				&  &200   &$0.968(0.032)$  &$0.650(0.350)$ &$0.937(0.063)$  &$0.000(1.000)$ &$0.000(1.000)$ &$0.927 (0.073)$ &$0.655 (0.345)$ &$0.030 (0.970)$\\
	\cmidrule(lr){2-18}
	 				&$t_3$
	 				&20   &$0.360(0.640) $ &$0.097( 0.903) $ &$0.307(0.693) $ &$0.001(0.999)$ &$0.000(1.000)$ &$0.263 (0.737)$ &$0.050 (0.950)$ &$0.479 (0.521)$\\
	 				&  &50    &$0.624 (0.376)$  &$0.223(0.777) $ &$0.551(0.449)$  &$0.000(1.000)$ &$0.000(1.000)$ &$0.515 (0.485)$ &$0.131 (0.869)$ &$0.025 (0.975)$\\
	 				&  &100    &$0.716 (0.284)$  &$0.277(0.723)$  &$0.631(0.369)$ &$0.000(1.000)$ &$0.000(1.000)$ &$0.569 (0.431)$ &$0.272 (0.728)$ &$0.000 (1.000)$\\
	 				&  &150    &$0.723 (0.277)$  &$0.270(0.730)$  & $0.640(0.360)$ &$0.001(0.999)$ &$0.000(1.000)$ &$0.584 (0.416)$ &$0.376 (0.624)$ &$0.000 (1.000)$\\
	 				&  &200    &$0.775 (0.225)$  &$0.268(0.732) $ &$0.703(0.297)$ &$0.000(1.000)$ &$0.000(1.000)$ &$0.642 (0.358)$ &$0.487 (0.513)$ &$0.000 (1.000)$\\
	 				\bottomrule[2pt]
	 		\end{tabular*}}
	 	\end{threeparttable}
	 	
	 \end{table}

Table \ref{tab:main3}  presents the frequencies of exact estimation and underestimation over 1000 replications under Setting A and Setting B. We set $k_{\max} = 10 $ for IterER,
$\alpha$-PCA-ER, Rit-ER, iTOP-IC/iTIP-IC, iTOP-ER/iTIP-ER. Under the normal case, the IterER and Rit-ER perform comparably, but both perform better than all other methods. The information-criterion methods iTOP-IC/iTIP-IC perform the worst in all cases.
As the tails of the idiosyncratic errors become heavier (from normal to $t_4,t_3$), all estimates deteriorate, especially for the $\alpha$-PCA-ER and TCorTh methods. The proposed Rit-ER method is the most robust and always performs the best for heavy-tailed data. As the sample size $T$ grows, the proportion of exact estimation by Rit-ER raises towards one.

\section{Real data example}

In this section, we illustrate the empirical performance of our proposed methods by analyzing a financial portfolio dataset, which was studied by \cite{wang2019factor} and \cite{Yu2021Projected}. The financial portfolio dataset is composed of monthly returns of 100 portfolios, well structured into a $10\times 10$ matrix at each time point,  with rows corresponding to 10 levels of market capital size (denoted as S1-S10) and columns corresponding to 10 levels of book-to-equity ratio (denoted as BE1-BE10). The dataset collects  monthly returns from January
1964 to December 2019 covering totally 672 months. The details are available at the website \url{http://mba.tuck.dartmouth.edu/pages} \url{/faculty/ken.french/data_library.html}.

\begin{figure}[!h]
  \centering
  \includegraphics[width=12cm]{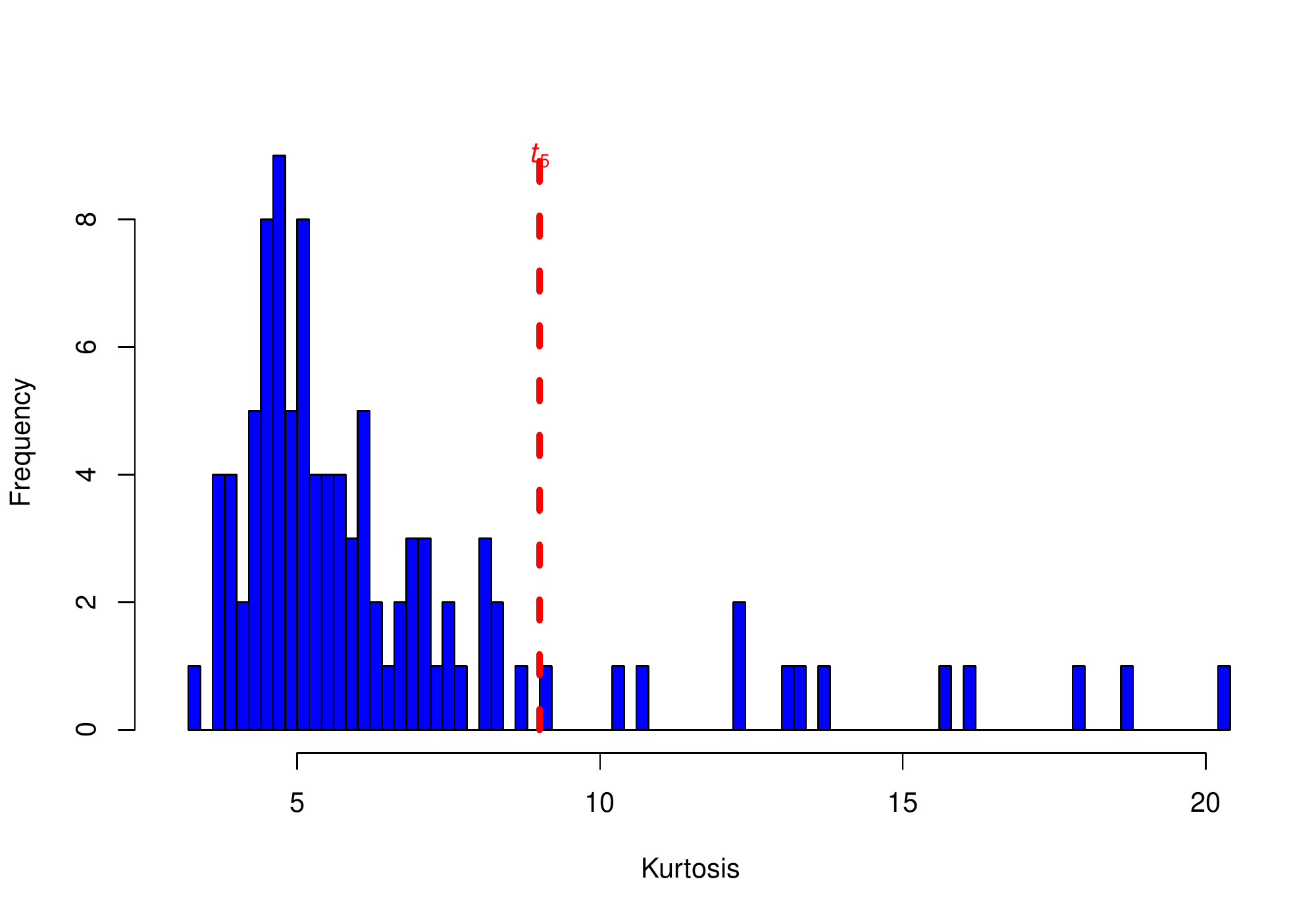}
 \caption{Histogram of the sample kurtosis for 100 portfolios and the red dashed line is the theoretical kurtosis of $t_5$ distribution.}
	\label{fig:kurtosis}
\end{figure}

Following the same preprocessing strategy by \cite{wang2019factor} and \cite{Yu2021Projected}, we first subtract monthly market
excess returns from the original return series and then standardize each of the series. We impute the missing values with the factor-model-based
method by \cite{xiong2019large}. The result of the augmented Dickey-Fuller test indicates the stationarity of the return series. The histogram of the sample kurtosis for 100 portfolios in Figure \ref{fig:kurtosis} demonstrates that the data are possibly heavy-tailed. The strong rule in \cite{He2021Vector} is implemented to diagnose the matrix factor structure in this dataset. For the parameters involved in the test, we set $\alpha=0.01, M=100$, $S\in\{200,300,400\}$, $f_1(S)=1-%
\protect\alpha-\protect\sqrt{2\ln S/S}$, $f_2(S)=1-\protect\alpha-S^{-1/3}$,
$f_3(S)=1-\protect\alpha-S^{-1/4}$, $f_4(S)=1-\protect\alpha-S^{-1/5}$, $%
f_5(S)=(1-\protect\alpha)/2$. The results for all test specifications show that there is overwhelming evidence in favour of
a matrix structure in the dataset.

 \begin{table*}[!h]
 	\begin{center}
 		\small
 		\addtolength{\tabcolsep}{1pt}
 		\caption{Loading matrices for Fama--French data set, after varimax rotation and scaling by 30.  ``RMFA": the  robust matrix factor analysis method, ``PE": the projected estimation method by \cite{Yu2021Projected},  $\alpha$-PCA: the method in \cite{fan2021} with $\alpha=0$, ``ACCE": the approach proposed by \cite{wang2019factor}, ``TOPUP" and ``TIPUP": the methods proposed by \cite{chen2022factor}, ``iTOPUP" and ``iTIPUP" : the methods proposed by \cite{Han2020Tensor}.
 		 }\label{tab6}
 		 \renewcommand{\arraystretch}{0.75}
 		\scalebox{1}{ 		 \begin{tabular*}{14.5cm}{cc|cccccccccc}
 				\toprule[1.2pt]
 				 \multicolumn{12}{l}{Size}\\
 				\toprule[1.2pt]
 				 Method&Factor&S1&S2&S3&S4&S5&S6&S7&S8&S9&S10\\ \hline
 				 \multirow{2}*{RMFA}&1&\cellcolor {Lavender}-16&\cellcolor {Lavender}-15&\cellcolor {Lavender}-13&\cellcolor {Lavender}-11&\cellcolor {Lavender}-8&-6&-3&0&4&\cellcolor {Lavender}6\\
 				 &2&\cellcolor {Lavender}-6&-2&2&5&\cellcolor {Lavender}8&\cellcolor {Lavender}10&\cellcolor {Lavender}12&\cellcolor {Lavender}14&\cellcolor {Lavender}15&\cellcolor {Lavender}10\\ \hline
 				 \multirow{2}*{PE}&1&\cellcolor {Lavender}-16&\cellcolor {Lavender}-15&\cellcolor {Lavender}-12&\cellcolor {Lavender}-10&\cellcolor {Lavender}-8&\cellcolor {Lavender}-5&-3&-1&4&\cellcolor {Lavender}7
 				\\
 				 &2&-6&-1&3&5&8&\cellcolor {Lavender}10&\cellcolor {Lavender}12&\cellcolor {Lavender}13&\cellcolor {Lavender}15&\cellcolor {Lavender}11
 				\\\hline
 	 			 \multirow{2}*{$\alpha$-PCA}&1&\cellcolor {Lavender}-14&\cellcolor {Lavender}-14&\cellcolor {Lavender}-13&\cellcolor {Lavender}-11&\cellcolor {Lavender}-9&\cellcolor {Lavender}-7&-4&-2&3&\cellcolor {Lavender}7
 	\\
 	&2&-4&-2&1&3&6&\cellcolor {Lavender}9&\cellcolor {Lavender}12&\cellcolor {Lavender}13&\cellcolor {Lavender}16&\cellcolor {Lavender}14
 	\\
 \hline
 	 			 \multirow{2}*{ACCE}&1&\cellcolor {Lavender}-12&\cellcolor {Lavender}-14&\cellcolor {Lavender}-12&\cellcolor {Lavender}-13&\cellcolor {Lavender}-10&\cellcolor {Lavender}-6&-3&-1&4&\cellcolor {Lavender}9
 	\\
 	&2&-1&-1&-1&2&5&\cellcolor {Lavender}10&\cellcolor {Lavender}11&\cellcolor {Lavender}18&\cellcolor {Lavender}15&\cellcolor {Lavender}11
 	\\ \hline
 	\multirow{2}*{TOPUP}&1&\cellcolor {Lavender}-12&\cellcolor {Lavender}-14&\cellcolor {Lavender}-12&\cellcolor {Lavender}-13&\cellcolor {Lavender}-10&\cellcolor {Lavender}-6&-3&-1&4&\cellcolor {Lavender}9\\
 	&2&-1&-1&-1&2&5&\cellcolor {Lavender}10&\cellcolor {Lavender}11&\cellcolor {Lavender}18&\cellcolor {Lavender}15&\cellcolor {Lavender}11\\\hline
 	\multirow{2}*{TIPUP}&1&0&0&0&-4&\cellcolor {Lavender}-8&\cellcolor {Lavender}-11&\cellcolor {Lavender}-10&\cellcolor {Lavender}-18&\cellcolor {Lavender}-13&\cellcolor {Lavender}-11
 	\\
 	&2&\cellcolor {Lavender}-13&\cellcolor {Lavender}-14&\cellcolor {Lavender}-13&\cellcolor {Lavender}-11&\cellcolor {Lavender}-7&\cellcolor {Lavender}-5&-4&1&5&\cellcolor {Lavender}11
 	\\\hline
 	\multirow{2}*{iTOPUP}&1&1&-1&1&-2&\cellcolor {Lavender}-6&\cellcolor {Lavender}-11&\cellcolor {Lavender}-10&\cellcolor {Lavender}-19&\cellcolor {Lavender}-13&\cellcolor {Lavender}-11
 	\\
 	&2&\cellcolor {Lavender}-13&\cellcolor {Lavender}-13&\cellcolor {Lavender}-12&\cellcolor {Lavender}-13&\cellcolor {Lavender}-10&\cellcolor {Lavender}-6&-3&0&4&\cellcolor {Lavender}10
 	\\
 	\hline
 	\multirow{2}*{iTIPUP}&1&1&0&1&-4&\cellcolor {Lavender}-7&\cellcolor {Lavender}-11&\cellcolor {Lavender}-11&\cellcolor {Lavender}-17&\cellcolor {Lavender}-13&\cellcolor {Lavender}-10
 	\\
 	&2&\cellcolor {Lavender}-14&\cellcolor {Lavender}-14&\cellcolor {Lavender}-13&\cellcolor {Lavender}-11&\cellcolor {Lavender}-8&\cellcolor {Lavender}-6&-2&0&\cellcolor {Lavender}5&\cellcolor {Lavender}9\\

 				\toprule[1.2pt]
\multicolumn{12}{l}{Book-to-Equity}\\
 				\toprule[1.2pt]
 				 Method&Factor&BE1&BE2&BE3&BE4&BE5&BE6&BE7&BE8&BE9&BE10\\\hline
 				 \multirow{2}*{RMFA}&1&\cellcolor {Lavender}6&1&-3&\cellcolor {Lavender}-6&\cellcolor {Lavender}-9&\cellcolor {Lavender}-11&\cellcolor {Lavender}-12&\cellcolor {Lavender}-13&\cellcolor {Lavender}-12&\cellcolor {Lavender}-11
 				\\
 				&2&\cellcolor {Lavender}19&\cellcolor {Lavender}17&\cellcolor {Lavender}12&\cellcolor {Lavender}9&5&3&0&-1&-1&0\\\hline
 				 \multirow{2}*{PE}&1&\cellcolor {Lavender}6&1&-4&\cellcolor {Lavender}-7&\cellcolor {Lavender}-10&\cellcolor {Lavender}-11&\cellcolor {Lavender}-12&\cellcolor {Lavender}-12&\cellcolor {Lavender}-12&\cellcolor {Lavender}-10
\\
&2&\cellcolor {Lavender}20&\cellcolor {Lavender}17&\cellcolor {Lavender}11&\cellcolor {Lavender}8&4&2&0&-1&-1&0
\\\hline
 				 \multirow{2}*{$\alpha$-PCA}&1&\cellcolor {Lavender}6&2&-4&\cellcolor {Lavender}-7&\cellcolor {Lavender}-10&\cellcolor {Lavender}-11&\cellcolor {Lavender}-12&\cellcolor {Lavender}-13&\cellcolor {Lavender}-12&\cellcolor {Lavender}-11
\\
&2&\cellcolor {Lavender}19&\cellcolor {Lavender}18&\cellcolor {Lavender}12&\cellcolor {Lavender}8&4&2&0&-1&-1&-1
\\\hline
 				 \multirow{2}*{ACCE}&1&\cellcolor {Lavender}6&-1&-4&\cellcolor {Lavender}-8&\cellcolor {Lavender}-8&\cellcolor {Lavender}-9&\cellcolor {Lavender}-10&\cellcolor {Lavender}-13&\cellcolor {Lavender}-15&\cellcolor {Lavender}-12
\\
&2&\cellcolor {Lavender}21&\cellcolor {Lavender}15&\cellcolor {Lavender}11&\cellcolor {Lavender}6&5&2&1&-2&-3&1
\\ 		 				\hline
\multirow{2}*{TOPUP}&1&\cellcolor {Lavender}6&-1&-4&\cellcolor {Lavender}-8&\cellcolor {Lavender}-8&\cellcolor {Lavender}-9&\cellcolor {Lavender}-10&\cellcolor {Lavender}-13&\cellcolor {Lavender}-15&\cellcolor {Lavender}-12
\\
&2&\cellcolor {Lavender}-21&\cellcolor {Lavender}-15&\cellcolor {Lavender}-11&\cellcolor {Lavender}-6&-5&-2&-1&2&3&-1\\\hline
\multirow{2}*{TIPUP}&1&\cellcolor {Lavender}-18&\cellcolor {Lavender}-14&\cellcolor {Lavender}-13&\cellcolor {Lavender}-10&\cellcolor {Lavender}-7&\cellcolor {Lavender}-6&-4&0&2&-1
\\
&2&\cellcolor {Lavender}-8&0&0&\cellcolor {Lavender}5&\cellcolor {Lavender}8&\cellcolor {Lavender}4&\cellcolor {Lavender}5&\cellcolor {Lavender}12&\cellcolor {Lavender}16&\cellcolor {Lavender}17
\\\hline
\multirow{2}*{iTOPUP}&1&\cellcolor {Lavender}6&0&-4&\cellcolor {Lavender}-8&\cellcolor {Lavender}-7&\cellcolor {Lavender}-9&\cellcolor {Lavender}-9&\cellcolor {Lavender}-13&\cellcolor {Lavender}-15&\cellcolor {Lavender}-13
\\
&2&\cellcolor {Lavender}-21&\cellcolor {Lavender}-16&\cellcolor {Lavender}-11&\cellcolor {Lavender}-7&\cellcolor {Lavender}-6&-3&-1&2&4&-3
\\\hline
\multirow{2}*{iTIPUP}&1&\cellcolor {Lavender}5&0&-1&\cellcolor {Lavender}-5&\cellcolor {Lavender}-9&\cellcolor {Lavender}-7&\cellcolor {Lavender}-7&\cellcolor {Lavender}-13&\cellcolor {Lavender}-15&\cellcolor {Lavender}-17
\\
&2&\cellcolor {Lavender}-17&\cellcolor {Lavender}-17&\cellcolor {Lavender}-12&\cellcolor {Lavender}-10&\cellcolor {Lavender}-7&-3&-3&3&4&0
\\ 		 				
 				\bottomrule[1.2pt]		
 		\end{tabular*}}		
 	\end{center}
 \end{table*}

The IterER method
suggests that $(k_1,k_2) = (2,1)$, the Rit-ER suggests that $(k_1,k_2) = (1,2)$, the $\alpha$-PCA-ER, iTOP-IC and iTIP-IC all suggest $(k_1,k_2) = (1,1)$, while iTOP-ER, iTIP-ER and TCorTh all suggest $(k_1,k_2) = (2,2)$.

\begin{table*}[!h]
	\begin{center}
		\small
		\addtolength{\tabcolsep}{0pt}
		\caption{Rolling validation for the Fama--French portfolios. $12n$ is the sample size of the training set. $k_1=k_2=k$ is the number of factors. $\overline{MSE}$, $\bar \rho$, $\bar v$ are the mean pricing error, mean unexplained proportion of total variances and mean variation of the estimated loading space. ``RMFA", ``PE'', ``ACCE'', ``$\alpha$-PCA'', ``TOPUP", ``TIPUP",  ``iTOPUP" and ``iTIPUP" are the same as in Table \ref{tab6}. }\label{tab8}
		 \renewcommand{\arraystretch}{0.4}
		\scalebox{1}{ 	
        \begin{tabular*}{13cm}{cc|cccccccc}
				\toprule[1.2pt]
				 $n$ &$k$ &RMFA&PE&$\alpha$-PCA&ACCE&TOPUP&TIPUP&iTOPUP&iTIPUP\\ \midrule[1.2pt]
				 &&\multicolumn{8}{c}{$\overline{MSE}$} \vspace{0.5em}
				  \\
				 5&1&{\bf0.860}&0.870&0.862&0.885&0.885&0.960&0.943&0.943\\
				 10&1&{\bf0.849}&0.855&0.860&0.880&0.880&0.950&0.937&0.943\\
				 15&1&{\bf0.850}&0.853&0.860&0.884&0.884&0.933&0.928&0.933
\\	 				
				 5&2&{\bf0.594}&0.596&0.601&0.667&0.667&0.704&0.706&0.703\\
				 10&2&{\bf0.600}&0.601&0.611&0.655&0.655&0.716&0.671&0.698\\
				 15&2&{\bf0.602}&0.603&0.612&0.637&0.637&0.695&0.655&0.686
\\	 				
				 5&3&{\bf0.518}&0.522&0.529&0.564&0.564&0.632&0.584&0.615\\
				 10&3&{\bf0.516}&0.519&0.526&0.573&0.573&0.616&0.583&0.600\\
				 15&3&{\bf0.515}&0.517&0.522&0.560&0.560&0.612&0.577&0.591\\
				\hline	
				&&\multicolumn{8}{c}{$\bar{\rho}$} \vspace{0.5em}\\
				 5&1&{\bf0.793}&0.802&0.796&0.828&0.828&0.943&0.933&0.935\\
				 10&1&{\bf0.776}&0.784&0.791&0.815&0.815&0.921&0.914&0.923\\
				 15&1&{\bf0.778}&0.782&0.792&0.812&0.812&0.890&0.907&0.914\\
	 				
				 5&2&{\bf0.618}&0.625&0.628&0.673&0.673&0.740&0.719&0.733\\
				 10&2&{\bf0.623}&0.628&0.636&0.668&0.668&0.738&0.687&0.728\\
				 15&2&{\bf0.622}&0.626&0.630&0.652&0.652&0.714&0.677&0.711\\
	 				
				 5&3&{\bf0.545}&0.550&0.556&0.590&0.590&0.664&0.610&0.637\\
				 10&3&{\bf0.544}&0.548&0.555&0.594&0.594&0.645&0.604&0.624\\				 15&3&{\bf0.541}&0.545&0.544&0.582&0.582&0.637&0.601&0.614\\
						\hline	
				&&\multicolumn{8}{c}{$\bar{v}$} \vspace{0.5em}\\
     			 5&1&{\bf0.144}&0.176&0.241&0.303&0.303&0.622&0.602&0.585\\					 10&1&{\bf0.062}&0.085&0.203&0.165&0.165&0.354&0.385&0.389\\
				 15&1&{\bf0.045}&0.064&0.234&0.152&0.152&0.291&0.347&0.345\\
	 				
				 5&2&{\bf0.163}&0.239&0.350&0.460&0.460&0.579&0.620&0.635\\
			    10&2&{\bf0.080}&0.092&0.261&0.257&0.257&0.427&0.419&0.429\\
			    15&2&{\bf0.050}&0.057&0.173&0.189&0.189&0.302&0.304&0.303\\
	 				
			    5&3&{\bf0.241}&0.286&0.432&0.497&0.497&0.628&0.556&0.599\\
			    10&3&{\bf0.110}&0.114&0.353&0.303&0.303&0.402&0.344&0.390\\
			    15&3&{\bf0.072}&0.084&0.308&0.299&0.299&0.384&0.330&0.353\\
				\bottomrule[1.2pt]		
		\end{tabular*}}		
	\end{center}
\end{table*}

The estimated row and column loading matrices after varimax rotation and scaling are reported in Table \ref{tab6}. From the table, we observe two different loading patterns across the rows (sizes). For the RMFA, PE, $\alpha$-PCA, ACCE, and TOPUP,
 the small size portfolios load heavily on the first factor while large size portfolios on the second factor, but for the TIPUP, iTOPUP and iTIPUP, the pattern reverses. We also find two distinct patterns across the columns (Book-to-Equity).

To further compare these methods, we use a rolling-validation procedure as in \cite{wang2019factor}. For each year $t$ from 1996 to 2019, we repeatedly use $n$ (bandwidth) years before $t$ to fit the matrix-variate factor model. The fitted loadings are then used to estimate the monthly factors and idiosyncratic errors in the current year $t$. Let $\Yb_t^i$ and $\hat\Yb_t^i$ be the observed and estimated price matrix of month $i$ in year $t$, and $\bar {\Yb}_t$ be the mean price matrix. Define
\[
\text{MSE}_t=\frac{1}{12\times10\times10}\sum_{i=1}^{12}\|\hat\Yb_t^i-\Yb_t^i\|_F^2, \quad \rho_t=\frac{\sum_{i=1}^{12}\|\hat\Yb_t^i-\Yb_t^i\|_F^2}{\sum_{i=1}^{12}\|\Yb_t^i-\bar\Yb_t\|_F^2},
\]
as the mean squared pricing error and unexplained proportion of total variances, respectively.  In  the rolling-validation procedure, the variation of loading space is measured by $v_t:=\mathcal{D}(\hat\Cb_t\otimes \hat\Rb_t,\hat\Cb_{t-1}\otimes \hat\Rb_{t-1})$.

We tried diverse combinations of $n$ and numbers of factors ($k_1=k_2=k$). We report the means of $MSE_t$, $\rho_t$ and $v_t$ in Table \ref{tab8}. The RMFA method has the lowest pricing errors across the board.

	\section{Discussion}
The current work studies the large-dimensional matrix factor model from the least squares and Huber Loss points of view. The KKT condition of  minimizing the residual sum of squares under the identifiability condition naturally motivates one to adopt the iterative projection estimation algorithm by \cite{Yu2021Projected}.  For the Huber loss, the corresponding KKT condition  motivates us to do PCA on weighted sample covariance matrix of the projected data.
We investigate the properties of theoretical minimizers of the squared loss and the Huber loss under the identifiability condition. We also propose robust estimators of the pair of factor numbers. The limitation of the current work lies in that we do not provide a theoretical guarantee of the estimators in the solution path of the iterative algorithm, which involves both statistical and computational accuracy.
We leave it as a future research direction.

\section*{Acknowledgements}

He's work is supported by  NSF China (12171282,11801316), National Statistical Scientific Research Key Project (2021LZ09), Young Scholars Program of Shandong University, Project funded by
China Postdoctoral Science Foundation (2021M701997) and the Fundamental Research Funds of Shandong University. Kong's work is partially supported by NSF China (71971118 and 11831008) and the WRJH-QNBJ Project and Qinglan Project of Jiangsu Province. Zhang's work is supported by  NSF China (11971116).


\bibliographystyle{model2-names}
\bibliography{ref}

\end{document}